\documentclass[fleqn,usenatbib]{mnras}

\usepackage{newtxtext,newtxmath}
\usepackage[T1]{fontenc}

\DeclareRobustCommand{\VAN}[3]{#2}
\let\VANthebibliography\thebibliography
\def\thebibliography{\DeclareRobustCommand{\VAN}[3]{##3}\VANthebibliography}

\usepackage{graphicx}
\usepackage{amsmath}

\newcommand{\rhowind}{\rho_{\mathrm{wind}}}
\newcommand{\rhocloud}{\rho_{\mathrm{cloud}}}
\newcommand{\rhomix}{\rho_{\mathrm{mix}}}
\newcommand{\Tcloud}{T_{\mathrm{cloud}}}
\newcommand{\Tmix}{T_{\mathrm{mix}}}
\newcommand{\Twind}{T_{\mathrm{wind}}}
\newcommand{\Mwind}{\mathcal{M}_{\mathrm{wind}}}

\newcommand{\nmix}{n_{\mathrm{mix}}}
\newcommand{\tcc}{t_{\mathrm{cc}}}
\newcommand{\tdrag}{t_{\mathrm{drag}}}
\newcommand{\tcool}{t_{\mathrm{cool}}}
\newcommand{\tcoolcloud}{t_{\mathrm{cool,cl}}}

\newcommand{\tcoolmix}{t_{\mathrm{cool,mix}}}
\newcommand{\cs}{c_{\mathrm{s}}}
\newcommand{\cscloud}{c_{\mathrm{s,cl}}}
\newcommand{\cswind}{c_{\mathrm{s,w}}}

\newcommand{\vwind}{v_{\mathrm{wind}}}

\newcommand{\rcloud}{r_{\mathrm{cloud}}}
\newcommand{\ratio}{\frac{\tcoolmix}{\tcc}}
\newcommand{\ratioinline}{\tcoolmix/\tcc}
\newcommand{\rcrit}{r_{\mathrm{crit}}}

\usepackage[utf8]{inputenc}
\usepackage{mathtools}
\usepackage{booktabs}
\usepackage{placeins}
\usepackage{physics}
\usepackage{orcidlink}
\usepackage{subcaption}
\usepackage{tabularx}

\title[Multi-cloud crushing]{Multi-cloud crushing -- the collective survival of cold clouds in galactic outflows}

\author[Seidl et al.]{ 
Benedikt S. Seidl,$^{\orcidlink{0009-0001-1731-8395}1, 2}$\thanks{E-mail: bseidl@mpa-garching.mpg.de}
Max Gronke,$^{\orcidlink{0000-0003-2491-060X}1}$
Ryan Jeffrey Farber,$^{\orcidlink{0000-0002-0649-9055}3,1}$
Klaus Dolag$^{2, 1}$
\\
$^{1}$Max Planck Institute for Astrophysics, Karl-Schwarzschild-Str. 1, D-85748 Garching, Germany\\
$^{2}$Universitäts-Sternwarte, Fakultät für Physik, Ludwig-Maximilians-Universität München, Scheinerstr. 1, D-81679 München, Germany\\
$^{3}$Department of Physics, Purdue University Fort Wayne, 2101 E. Coliseum Blvd, Fort Wayne, IN 46805, USA
}

\date{Accepted XXX. Received YYY; in original form ZZZ}

\pubyear{2025}

\begin{document}
\label{firstpage}
\pagerange{\pageref{firstpage}--\pageref{lastpage}}
\maketitle

\begin{abstract}
The ram-pressure acceleration of cold gas by hot outflows plays a crucial role in the dynamics of multiphase galactic winds.
Recent numerical studies incorporating radiative cooling have identified a size threshold for idealized cold clouds to survive within high-velocity outflows. 
This study extends the investigation to a more complex morphology of cold gas as observed in the interstellar medium.
We conduct three-dimensional hydrodynamic simulations of ensembles of individual spherical clouds to systematically explore under which conditions the cold clouds can survive. 
We find that cloud ensembles can survive collectively -- even when individual clouds, if isolated, would be rapidly destroyed. 
Our results indicate that, besides the morphology, factors such as tight packing, small inter-cloud distance and higher fragmentation facilitate survival.
We propose a novel multi-cloud survival criterion that accounts for collective properties of the cloud system, including total gas mass and the geometric configuration based on an effective volume filling fraction of the cold gas $F_V$.
This fraction is computed by constructing a composite volume from individual enclosing conical boxes aligned with the wind, incorporating spatial overlap and cloud-tail spreading.
The box dimensions scale with the critical survival radius $r_{\rm crit}$ from the single-cloud criterion. 
We find a universal threshold $F_{V,{\rm crit}}\approx 0.24$ that robustly separates surviving from destroyed systems across diverse geometric configurations.
Our findings emphasize the critical importance of initial cloud distribution and fragmentation in governing the long-term evolution and survival of cold gas structures, providing insight into observed multiphase outflows and CGM dynamics.
\end{abstract}



\section{Introduction} \label{sec:intro}
In dynamical models of galaxy evolution, gas inflows and outflows are regulated by feedback processes, including those driven by active galactic nuclei (AGN) and stars \citep[]{somerville2015physical, cgm_review, fauchergiguere2023key}.
The continuous exchange of gas between galaxies and their CGM, often described as the galactic baryon cycle or `galactic ecosystem', plays a crucial role in understanding how galaxies can form stars much longer than their local gas depletion times \citep{bigiel2011constant,semenov2017physical}. 
Moreover, the CGM contains large amounts of baryonic matter and acts as an important gas reservoir, feeding star formation and influencing observable galaxy properties (see \citealp{cgm_review, Peroux_Howk_2020, Knapen_Outskirts} for detailed reviews). 

Thus, understanding the CGM and the galactic baryon cycle plays an essential role moving towards a comprehensive picture of galaxy formation and evolution. 
Studies using absorption lines in background quasi-stellar objects (QSOs) (e.g., \citealp{Rauch_1998, Weng_2023}) and direct observations of emission halos (e.g., \citealp{Cantalupo_etal_2014, Battaia_Hennawi_Prochaska_Onorbe_Farina_Cantalupo_Lusso_2019}) have shed light on the multiphase physical state of the CGM; that is, gas phases of vastly different temperatures are located co-spatially.

Specifically, the \textit{cold gas} (with temperatures $\lesssim10^4\,\mathrm{K}$) is predominantly neutral, containing low ions and dust.
The \textit{warm} (alternatively \textit{intermediate temperature} or \textit{mixed temperature}) \textit{gas} spans temperatures from $10^4$ to $10^6\,\mathrm{K}$.
Finally, the hot gas (with $T \gtrsim 10^6\,\mathrm{K}$) is approximately the virial temperatures of galactic halos in the CGM, or orders of magnitude hotter in galactic winds; therefore, the hot gas is rarely detectable.

Absorption features from low-ionization species (e.g., Hydrogen Ly$\alpha$, Mg~II, C~II, Si~II) and high-ionization species (e.g., C~IV, O~VI, N~V, O~VII) in quasar spectra provide evidence for these distinct phases \citep[see][and references therein for further observational footprints of multiphase gas]{cgm_review}.
While the cold and warm phases are thought to dominate a galaxy’s baryon mass, the mass fraction of the hot phase remains uncertain due to observational challenges \citep{Rupke_2018,qu2020warm,Veilleux_Maiolino_Bolatto_Aalto_2020}, especially since the hot phase is predominantly observable via X-rays.

In many cases, only photometric data from the cold gas component is available, despite the fact that mechanisms like ram-pressure acceleration can generate observable spectroscopic signatures (e.g., velocity fields; see \citealt{Veilleux_Cecil_Bland-Hawthorn_2005}). 
Moreover, the close interplay between cold gas formation and star formation processes \citep{Zhang_2018} underscores the importance of accurately modeling the multiphase nature of galactic winds.

On the theoretical side, cosmological simulations have enabled increasingly realistic representations of the cosmos.
Yet, such simulations still lack the resolution necessary to capture the multiphase morphology and dynamics of galactic winds and cosmic ecosystems. Thus, important physical processes occurring on small scales in multiphase systems (e.g., radiative cooling) remain unresolved and can lead to non-convergent results \citep[e.g.,][]{Hummels_Smith_Hopkins_O’Shea_Silvia_Werk_Lehner_Wise_Collins_Butsky_2019,peeples2019figuring,van2019cosmological}.
This motivates the need for targeted small-scale simulations to understand how resolving cold gas systems may refine our models of galaxy formation.

The general setup of cloud-crushing problems was established circa 50 years ago \citep{McKee_Cowie_1975,McKee_Cowie_Ostriker_1978,Klein_McKee_Colella_1994}.
A cold and dense, spherical cloud of gas is exposed to a hot and tenuous, often supersonic, wind. 
On impact, a Kelvin-Helmholtz instability arises at the outer borders of the cloud where the shear velocity with respect to the hot gas is high. 
Additionally, a Rayleigh-Taylor instability forms, when the cold, dense medium is accelerated by the wind.
Many numerical studies demonstrated cold clouds are quickly destroyed by mixing from these instabilities \citep{Pittard_Dyson_Falle_Hartquist_2005,Scannapieco_Brueggen_2015,Schneider_Robertson_2017}.
Yet, observational evidence, as introduced earlier, shows an abundance of fast moving cold and warm gas, demanding cloud survival and possibly cloud growth through entrainment \citep{Veilleux_Cecil_Bland-Hawthorn_2005}.

Before the late 2010s, most studies on cloud–wind interactions did not fully capture the phenomenon of cloud entrainment.
While several works incorporated radiative cooling in simulations \citep{Mellema_Kurk_Roettgering_2002, Scannapieco_Brueggen_2015, Schneider_Robertson_2017}, they generally found rapid cloud destruction rather than sustained survival or growth during their simulation runtime. 
However, more recent studies have highlighted the crucial role of turbulent mixing layers in mediating interactions between the cold cloud and the surrounding hot wind that enable the growth of clouds at longer timescales. 
These mixing layers facilitate the entrainment of hot gas, which, when combined with efficient cooling, can lead to cloud stabilization and even growth \citep[e.g.,][]{armillotta2017survival, Gronke_2018, Li_Hopkins_Squire_Hummels_2020}. 
This growing body of work suggests that cold gas in galactic winds can be much more resilient than previously thought, with implications for the persistence and recycling of multiphase material in galaxy evolution.

As turbulent mixing at the cloud–wind interface produces warm gas, the enhanced cooling in this region facilitates the growth of the cold phase.
The schematic representation of a turbulent mixing layer (TML) is provided in \autoref{fig:ccschema}.
The mixed gas has a temperature at the geometric mean of the hot and cold temperatures \citep{Begelman_Fabian_1990}, at which cooling is markedly more efficient.
As the hot gas impinges upon the colder, denser gas, Kelvin–Helmholtz instabilities develop, rapidly mixing the two phases and forming a reservoir of warm, rapidly cooling gas.
This process not only augments the cold gas mass but also transfers significant momentum to it, thereby stabilizing the cloud’s tail and promoting its entrainment with the wind \citep{Gronke_2018,Gronke_Oh_2020,Tonnesen2021}.

While previous `cloud crushing' simulations investigated the effect of a hot wind on a single isolated cloud, realistic multiphase media in the interstellar medium (ISM) and galactic winds are more complex. 
In particular, even if individual clouds can be characterized in galactic winds \citep[e.g.][]{di2019molecular, di2020cold, lopez2025observational}, they might be located close enough to each other to affect their dynamics and ultimate evolution.
Specifically, `shielding' of different clouds can be an important effect which has been discussed with adiabatic simulations in the literature (e.g., \citealp{villares2024hydrodynamic} and references therein; see below) --  an effect which can be stronger with cooling due to the multiple sources of mixed gas.

In this paper, we focus on the interaction between hot, supersonic winds and cold gas cloud ensembles, expected to exist in the ISM and CGM.
Employing high-resolution hydrodynamical simulations, we extend earlier studies that considered a single spherical cloud \citep{Klein_McKee_Colella_1994, McKee_Cowie_1975, Gronke_2018} by exploring more complex, non-spherical configurations composed of multiple spherical clouds. 
This approach permits us both to test the classical survival criterion on individual clouds and to investigate the effects of realistic cloud morphology on survival and entrainment.

There have been several previous attempts to investigate multi-cloud setups, although most of these studies do not incorporate radiative cooling in the same manner as presented here.
Early investigations by \citet{Cowie_McKee_Ostriker_1981} and \citet{Poludnenko_Frank_Blackman_2002} explored two-dimensional versions of the problem.
While \citet{Cowie_McKee_Ostriker_1981} considered a very general scenario, \citet{Poludnenko_Frank_Blackman_2002} focused on identifying a critical cloud separation necessary for survival in hot outflows.
In a similar spirit to our methodology, \citet{Aluzas_Pittard_Hartquist_Falle_Langton_2012} examined individual spherical clouds; however, their study centered on the evolution of post-shock turbulence rather than the radiative cooling processes that are central to our work.
The same group extended their investigation in \citet{Aluzas_Pittard_Falle_Hartquist_2014} by introducing magnetic fields, a factor that is not considered in this project.
Moreover, the studies by \citet{Banda-Barragan_Brueggen_Federrath_Wagner_Scannapieco_Cottle_2020} and \citet{Banda-Barragan_Brueggen_Heesen_Scannapieco_Cottle_Federrath_Wagner_2021} offer an interesting perspective by employing a more realistic density distribution alongside cooling effects, and their findings on shock evolution in the mixed medium are consistent with earlier work.
Additionally, one of the configurations explored here has also been addressed by \citet{Forbes_Lin_2019} in a cooling setup aimed at determining a critical cloud distance. 

Our project seeks to build on these previous efforts by systematically studying a variety of cloud ensemble morphologies with radiative cooling, developing a generalized survival criterion that better reflects the diversity of structures observed in the CGM and galactic outflows.

The outline of this paper is as follows. 
In \autoref{sec:methods} we introduce the quantities describing the classical cloud-crushing problem and put our system into physical scale. 
Additionally, we review the status quo for a survival criterion and describe the numerical methods employed to fully capture the problem.
In \autoref{subsec:morphology} we propose several systems of multi-cloud morphology with an ad-hoc guess for an adjusted survival criterion.
\autoref{sec:results} covers the result of the previously introduced setups in detail, focusing on mass growth and diagrams for varying cloud parameters in different initial setups.
In \autoref{sec:discussion} we propose a new composite survival criterion aiming to capture the variety of setups probed before summarizing and concluding in \autoref{sec:conclusion}.   

\section{Methods} \label{sec:methods}
\subsection{Analytical Considerations} \label{subsec:analytical}
\begin{figure}
    \centering
    \includegraphics[width=\columnwidth]{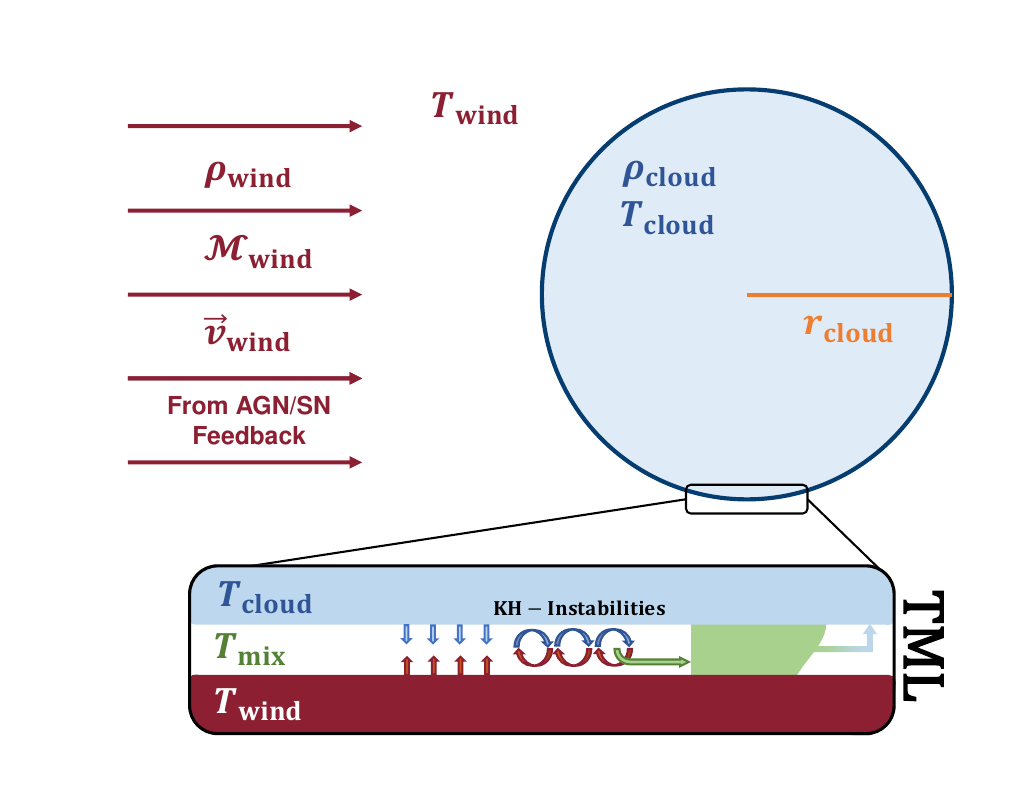}
    \caption{A setup of the cloud-crushing problem and an overview of common parameters.
The zoom-in schematically introduces the physics of a turbulent mixing layer, the ultimate driver of cloud growth. Colored red, green, and blue are the hot, mixed, and cold gas respectively, where the mixed gas of intermediate temperature quickly cools and thus feeds the cold gas reservoir.}
    \label{fig:ccschema}
\end{figure}

Most of the key parameters were defined by \citet{Klein_McKee_Colella_1994}.
\autoref{fig:ccschema} shows a visual representation of the key parameters for this problem.

Central to the problem is the cloud overdensity $\chi = \rhocloud/\rhowind= \Twind/\Tcloud$ where the last equality holds for pressure equilibrium.
Due to the self-similar nature of the problem without cooling, we use a natural unit system where $k_B = 1$.
We denote the speed of sound for the hot (wind) and cold (cloud) phase as \begin{align}\label{eq:cswind-cscloud}
    \cswind &=  \left(\gamma \Twind\right)^{1/2} \, ,\\
    \cscloud &= \left(\gamma \Tcloud\right)^{1/2} = \left(\gamma \Twind \chi^{-1}\right)^{1/2} = \cswind \chi^{-1/2} \, .
\end{align}

The Mach number of the hot wind is thus $\Mwind = \vwind/\cswind$ or correspondingly the wind velocity can be written as $\vwind = \Mwind \left(\gamma \Tcloud\right)^{1/2} \chi^{1/2}$.

Essential to this problem is the mixing (and consequential potential cooling) of the gas. 
\citet{Begelman_Fabian_1990} provide a simple model of turbulent mixing layers driven by Kelvin-Helmholtz instabilities.
Under the assumption that the efficiency factors, defined as how much of their mass and energy the two phases put into the mixing layer, are of equal magnitude, they find that the temperature in the mixing layer is the geometric mean of the two temperatures, i.e.,
\begin{equation} \label{eq:tmix}
    \Tmix = \left(\Tcloud \Twind\right) ^{1/2} = \Tcloud \chi^{1/2}.
\end{equation}
From the pressure equilibrium of all phases, we additionally find the mixing density of the gas in the TML $\rhomix = \rhocloud  \Tcloud \Tmix^{-1}= \rhocloud \chi^{-1/2} =\rhowind  \chi^{1/2}$.
Since the temperatures of the cold and hot phases differ by orders of magnitude, this result imposes a widely different mixing temperature than the arithmetic mean. 
In particular, the geometric mean temperature implies rapid cooling in the mixing layer due to the shape of the cooling function \citep[e.g.][]{Sutherland_Dopita_1993}.

\subsubsection{Timescales}
\citet{Klein_McKee_Colella_1994} derive the \textit{cloud-crushing time}, which is essential for classifying cloud survival and growth; we expect the cloud to be destroyed within a few cloud-crushing times if we do not consider any cooling effects.
The cloud crushing time is approximately equal or equivalent to \citep[][]{Chandrasekhar_1961} the Kelvin-Helmholtz timescale of the cloud experiencing shear, the Rayleigh Taylor timescale of the cloud being accelerated through ram pressure, or the \textit{sound-crossing time} of the cloud $t_{\mathrm{sc}} = \rcloud/\cs$ -- and is thus given by
\begin{equation} \label{eq:tcc}
    \tcc \equiv \chi^{1/2}\frac{\rcloud}{\vwind} = \Mwind^{-1} t_{\mathrm{sc}} \, .
\end{equation}

The Kelvin-Helmholtz timescale for a given wave number of the instability $k$ is defined as $ t_{\mathrm{KH}}^{-1} = k \chi^{-1/2} \vwind$.
\citet{Klein_McKee_Colella_1994} set the Rayleigh-Taylor acceleration, that appears in Chandrasekhar's approach, to $a = \rcloud/\tcc^2$ which yields $t_{\mathrm{RT}}^{-1} = \left(\rcloud k\right)^{1/2}\tcc^{-1}$.
Assuming that $k \rcloud \approx 1$ given that the instability arises on a typical scale of one cloud radius, both timescales are within order unity of the cloud crushing time \citep[][Sections X and XI]{Chandrasekhar_1961}.

The central problem to cloud crushing becomes apparent when comparing the cloud-crushing timescale to the drag timescale $\tdrag = \chi \rcloud/\vwind=\chi^{1/2} \tcc$; the drag timescale arises after simple considerations of how much momentum the wind must transfer into the cloud's momentum to make it co-moving.
With typical values of $\chi \sim 100 - 1000$ the additional factor of $\sqrt{\chi}$ in $\tdrag$ causes the cloud crushing time to be $\sim 10 - 30$ times smaller. 
Therefore, a cloud will be destroyed before it can be dragged to a co-moving velocity with the hot wind \citep[see][for a observationally supported summary of this `entrainment problem']{Zhang_Thompson_Quataert_Murray_2017}.

With radiative cooling, the last timescale we need to introduce is the cooling time using the thermal energy density per unit volume in its most basic definition
\begin{equation}\label{eq:tcool}
    \tcool = \frac{E}{\dot E} = \frac{3nkT}{2n^2\Lambda(T)} \sim \frac{kT}{n \Lambda(T)}.
\end{equation}
Of particular interest is the cooling time in the mixing layer defined as
\begin{align}
    \tcoolmix &= \frac{k\Tmix}{\nmix \Lambda(\Tmix)}= \chi \frac{\Lambda(\Tcloud)}{\Lambda(\Tmix)} \tcoolcloud. \label{eq:tcoolmix}
\end{align}
\autoref{tab:timescale_overview} summarizes the timescales for later reference.

\begin{table}
    \centering
    \caption{Overview table of the various timescales used in the cloud crushing problem.}
    \def\arraystretch{1.3}
    \begin{tabular}{lll}\toprule
         Timescale & Parameter &  Definition\\ \midrule
         Sound crossing time &$t_{\mathrm{sc}}$&  $\rcloud\cs^{-1}$ \\
            Drag time   & $\tdrag$& $\chi \rcloud\vwind^{-1}= \chi^{1/2} \tcc$\\
            Cloud crushing time &$\tcc$& $\chi^{1/2}\rcloud\vwind^{-1} = \Mwind^{-1}  t_{\mathrm{sc}}$\\
            KH timescale &$t_{\mathrm{KH}}^{-1} $&$ k \chi^{-1/2} \vwind$ \\
            RT timescale &$t_{\mathrm{RT}}^{-1}$&$ \left(\rcloud k\right)^{1/2} \tcc^{-1}$ \\
            Cooling time &$\tcool$ & $ kT\left(n \Lambda(T)\right)^{-1}$ \\
            \bottomrule
            \end{tabular}
    \label{tab:timescale_overview}
\end{table}

\subsection{A Survival Criterion for Cold Clouds}
The basic idea behind cloud survival undergoing ram pressure acceleration (and cooling), is that if the system can cool down mixed gas faster than it gets destroyed, we observe survival.
The cloud can therefore restock its cold gas contents faster than it loses that gas due to the effects of shocks, turbulence, and shattering.
This survival criterion can be put in the form \citep{Gronke_2018}
\begin{equation}
    \ratio < 1
    \label{eq:survivalcriterion}
\end{equation}
where two previously defined timescales come into play.
We note that since \citet{Gronke_2018}, other alternative survival criteria have been suggested and compared to each other \citep[e.g.][]{Li_Hopkins_Squire_Hummels_2020,sparre2020interaction, kanjilal2021growth,Farber_Gronke_2022,Abruzzo_Fielding_Bryan_2023}, however, this work is not about the exact form of the survival criterion but instead aims to loosen the assumption of an individual cloud undergoing ram pressure acceleration.

It might be noteworthy that a timescale criterion of the form in \autoref{eq:survivalcriterion} can be rewritten to $r_{\rm cl} > r_{\rm cl, crit}$, i.e., clouds larger than a certain critical radius will survive. For the specific choice of $\ratio < 1$, this yields

\begin{equation}
        r_{\mathrm{cl,crit}} \approx 1.7 \,\mathrm{pc}\frac{T_{\mathrm{cloud,4}}^{5/2} \Mwind}{P_3 \Lambda_{\mathrm{mix, -21.4}}} \frac{\chi}{100}
\end{equation}
\noindent where  $T_{\mathrm{cloud,4}} \equiv \Tcloud/10^4\mathrm{K}$, $P_3 \equiv nkT/10^3\mathrm{cm}^{-3}\mathrm{K}$, and $\Lambda_{\mathrm{mix,-21.4}} \equiv\Lambda(\Tmix)/10^{-21.4}\mathrm{erg}\,\mathrm{cm}^3\mathrm{s}^{-1}$.

\begin{figure*}
    \centering
    \includegraphics[width=\textwidth]{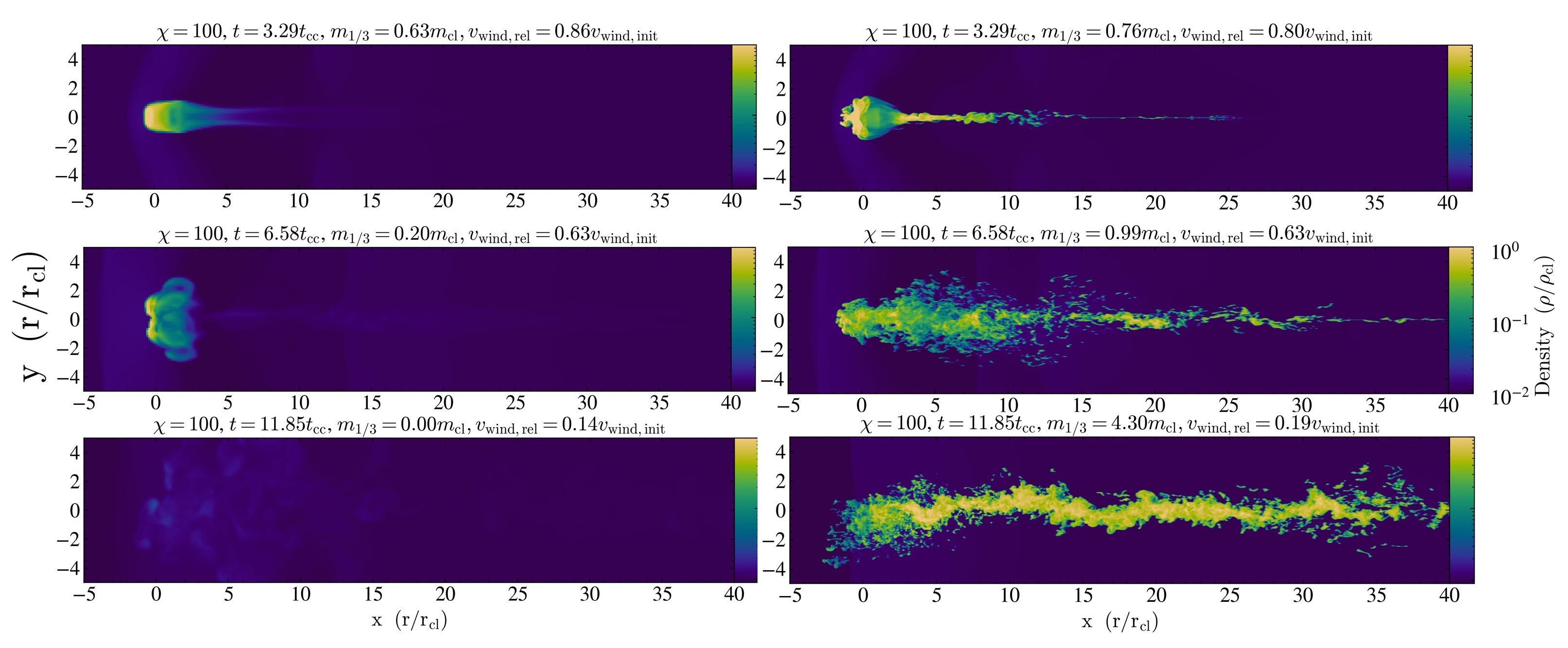}
    \caption{The evolution of a cloud for survival and destruction.
The timestamps are identical for both cases.
The left-hand side shows 8 cells per $\rcloud$ resolution, and destruction of a cloud with $\ratioinline \approx 7.55$.
The right-hand side shows a 32 cells per $\rcloud$ resolution survival of a cloud with $\ratioinline \approx 7.55\times 10^{-2}$.
Visualized is a mass-weighted projection plot of the density in the simulation box. Indicated above the snapshots one finds several measurable quantities, indicating the gas mass and relative velocity in the box. Entrainment allows the cloud to surpass the initial mass after several $\tcc$.}
    \label{fig:paper_replica_snapshots}
\end{figure*}

The evolution of the cloud can differ drastically depending on $\ratioinline$.
\autoref{fig:paper_replica_snapshots} shows two simulations where $\ratioinline$ varies by two orders of magnitude.

\subsection{Numerical Methods} \label{subsec:nummethods}
We simulate cloud–wind interactions using Athena 4.2 on a Cartesian, uniform grid \citep{Stone_2008}. The code employs second-order reconstruction with limiting in the characteristic variables, an HLLC Riemann solver, and a 3D van Leer unsplit integrator. In addition to standard hydrodynamic variables, we introduce scalar tracers to track the cloud material in each grid cell.

Cooling is implemented following \citet{Townsend_2009} with a seven-piece power-law fit to the \citet{Sutherland_Dopita_1993} cooling curves at solar metallicity $Z=Z_{\sun}$. Scaling factors, as well as imposed floors ($10^4\,\mathrm{K}$) and ceilings ($0.6\,\Twind$) on temperature and density (floor), ensure realistic cloud survival and maintain self-similar mass growth.
To distinguish cloud gas from wind gas, in a process that is known as \textit{entrainment}, we declare any mass of density higher than $1/3$ of the cloud's initial density to be cold gas mass. If this combined gas mass does not vanish over time (usually a threshold of one percent of the initial mass) we call it entrainment, and where it surpasses the initial cloud's mass, we call it \textit{cloud growth}.

We set the cloud radius aiming for a resolution of 8 cells per radius. Box dimensions follow the criteria suggested by \citet{Gronke_2018}, modified for a multi-cloud configuration. The box usually extends from $0.5\chi\rcloud$ to $\chi\rcloud$ downstream of the most distant cloud, and its lateral width -- ranging between 5 and 20 $\rcloud$ -- ensures that even cold gas ejected far from the primary tail is captured (see \autoref{app:parameters}).

Since the wind drags the cloud over distances that can exceed 100 cloud radii (given by $x_{\mathrm{travel}}\sim\vwind\tdrag=\chi\rcloud$), we dynamically shift the simulation box. As in \citet{Shin_Stone_Snyder_2008}, \citet{McCourt_OLeary_Madigan_Quataert_2015}, and \citet{Gronke_2018} (detailed in \citealt{Dutta_Sharma_2019}), tracer scalars allow us to track the cloud’s original material. The average velocity in the wind direction is computed via
\begin{equation}
    \langle v_{\mathrm{x}} \rangle = \frac{\int \rho C v_{\mathrm{x}} dV}{\int \rho C dV}
\end{equation}
and the kinetic energy and momentum in each cell are adjusted accordingly. This effectively shifts the frame so that the simulation becomes co-moving with the cloud, ensuring the cloud remains within the computational domain.

\begin{figure}
    \centering
    \includegraphics[width=\columnwidth]{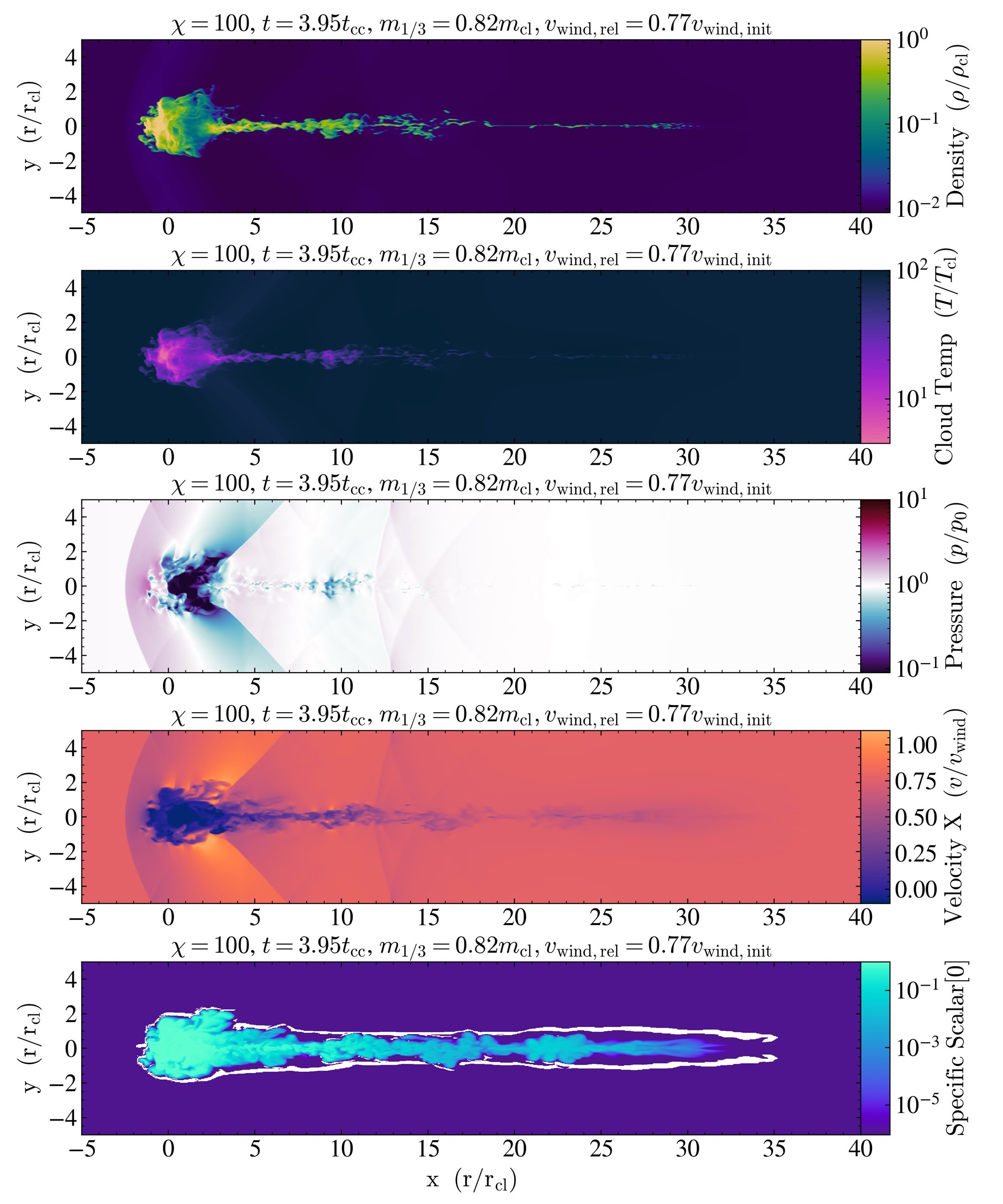}
    \caption{A visualization of different properties introduced in the problem.
Most plots are mass-weighted means, except the pressure and specific scalar (tracer) plots.
Most notably, one observes several shock-fronts throughout the cloud and its surrounding
The data is taken from a high-resolution survival run.}
    \label{fig:plotstyles}
\end{figure}

We visualize the simulations mostly via the mass-weighted mean of the density field.
\autoref{fig:plotstyles} shows a variety of other quantities and once again the geometry of the cloud crushing problems and its shock fronts.

\subsection{Multi-cloud morphology} \label{subsec:morphology}
In our multi-cloud simulations, several parameters are held fixed.
The cold gas temperature is set to \(\Tcloud = 10^4\,\mathrm{K}\) with an overdensity \(\chi=100\), and unless noted otherwise, the ratio $\ratioinline$ is fixed at approximately 1.03 (its specific choice is discussed below). 
The wind velocity is chosen to yield a Mach number \(\Mwind = 1.5\).
We impose an absolute temperature floor \(T_{\mathrm{floor}}=10^4\,\mathrm{K}\) and an upper limit of \(10^7\,\mathrm{K}\), while radiative cooling is suppressed above \(T_{\mathrm{ceil,cool}}=0.6\,\Twind=6\times10^5\,\mathrm{K}\) (only some early runs deviate from this with a temperature ceiling of $\Twind$, which does not affect the outcome). 
All initial conditions use a resolution of 8 cells per cloud radius, and density plots are mass-weighted.

To justify our choice of \(\ratioinline \approx1.03\), we first consider survival in a multi-cloud configuration by examining three basic arrangements: a \textit{wind-aligned} setup (clouds arranged along the wind direction), an \textit{orthogonal} setup (clouds arranged perpendicular to the wind), and intermediate ellipsoidal or spherical configurations.

We represent the overall structure as an ensemble of smaller spherical clouds, thereby emulating a larger object (e.g., a sphere or slab) while preserving the single-cloud survival criteria and related timescales.
Throughout this study, quantities such as \(\ratioinline\) and cloud size \(\rcloud\) refer exclusively to individual clouds, with temperature and overdensity parameters remaining consistent with the single-cloud scenario.

\subsubsection{Effective Radius} \label{subsubsec:effective_radius}
By summing the masses of the individual clouds, we define an \textit{effective radius} for a composite cloud with a single cloud density $\rhocloud$, enabling the application of the single-cloud survival theory.
Under constant wind velocity, overdensity, and cold-gas temperature, the effective \(\ratioinline\) is reduced -- as dictated by \autoref{eq:tcc} -- since increasing mass enlarges the cloud radius and lowers the ratio. 
We define the effective radius for $N_{\rm cl}$ clouds as
\begin{equation} \label{eq:effradius}
    r_\mathrm{cl,eff} = N_\mathrm{cl}^{1/3} \rcloud,
\end{equation}
i.e., an enclosing sphere with radius $r_{\rm cl,eff}$ and density $\rhocloud$ contains the same mass as all clouds combined.
For fixed $r_{\rm cl}<r_{\rm cl,crit}$, this implies a threshold for the number of clouds we need to place in the system to fulfill the survival criterion for the effective radius $r_\mathrm{cl,eff}\geq \rcrit$.

Furthermore, we define a \textit{mass limit} that is equivalent to the effective radius, which describes the mass of the cloud with $r_\mathrm{cl, eff}$ in radius and the cold gas density $\rhocloud$.
For instance, scaling the cloud radius by a factor $\alpha$, to reach the effective radius, changes the total cold gas mass as $m^\prime = \frac{4}{3}\pi \rhocloud(r_\mathrm{cl,eff})^3 = N_\mathrm{cl}m_{\rm cl} =\alpha^3 m$.    

\begin{figure}
    \centering
    \includegraphics[width=\columnwidth]{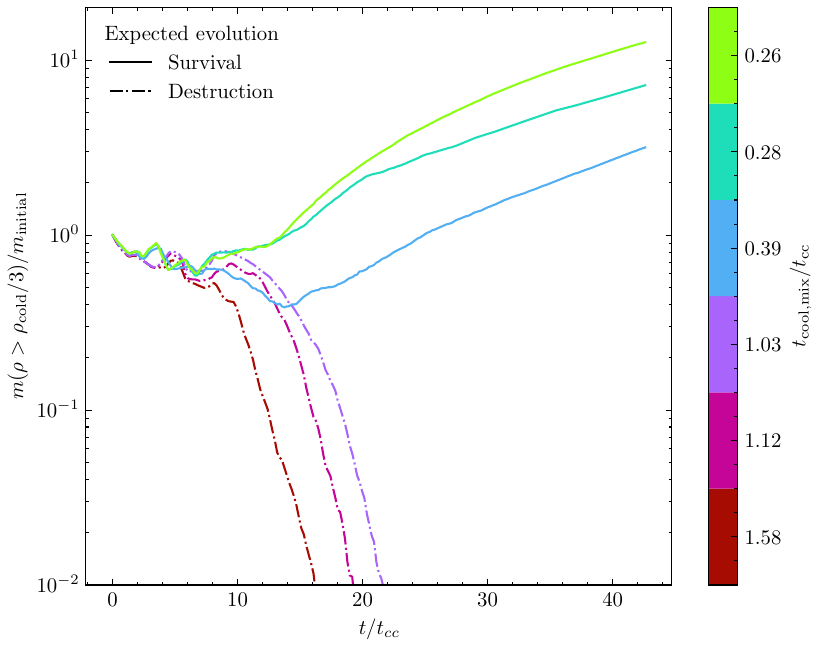}
    \caption{Study of the transitional regime between survival and destruction at different ratios of $ t_{\mathrm{cool,mix}}/t_{\mathrm{cc}}$. Shown for different values of $\protect t_{\mathrm{cool,mix}}/t_{\mathrm{cc}}$ is the cold gas mass fraction up to 40 cloud crushing times.
The simulations were carried out with a single cloud to study behavior close to the survival criterion, whose expected outcome is denoted by two different linestyles.
}
    \label{fig:growth_study}
\end{figure}

Our first step in the new cloud-crushing simulations was to identify suitable values of $\ratioinline$ close to the survival threshold such that the number of clouds in the simulations remains practical.
\autoref{fig:growth_study} shows the normalized cold gas mass versus time (in units of the cloud-crushing time) for various \(\ratioinline\) values near the survival threshold (with survival expected for \(\ratioinline<1\) and destruction for \(\ratioinline>1\) ).
The mass evolutions shown in \autoref{fig:growth_study} indicate behavior consistent with the survival criterion (Eq.~\eqref{eq:survivalcriterion}).
The governing parameters investigated in the following sections are the edge-to-edge distance between neighboring clouds (inter-cloud separation), the number of clouds and the shape of the initial configuration.
This justifies our fiducial choice of $r_{\rm cl,crit}/r_{\rm cl}=\ratioinline\approx 1.03$ as this requires a `rescaling factor' of the cloud's radius of approximately $\alpha=N_\mathrm{cl}^{1/3}=2.6$ in order to achieve  $r_{\rm cl,crit}/r_{\rm cl,eff}=\ratioinline\approx 0.39$ (see \autoref{fig:growth_study} and the certain survival for a single cloud of that size) where $r_{\rm cl,eff}/r_{\rm cl,crit}\approx 2.56 > 1$ by putting $N_\mathrm{cl} \approx 18$ small clouds into the box, making it computationally feasible.
While larger values of $r_{\rm cl, crit}/r_{\rm cl}=\ratioinline$ for the single clouds would also be an interesting case, the resolution requirement of at least 8 grid cells per cloud radius (cf. \S~\ref{subsec:nummethods}) and the condition $r_{\rm cl,eff}>r_{\rm cl,crit}$ requires computational resources beyond the scope of this study. We thus defer this to future work.
\section{Results} \label{sec:results}
In this section, we provide a detailed analysis of the numerical simulations.
Specifically, we present two limiting cases. 
Firstly, a setup where the clouds are entirely aligned with the wind and placed next to each other along the wind direction (\autoref{res:aligned}).
Secondly, a setup where clouds are placed side-by-side perpendicular to the wind direction (\autoref{res:orthogonal}), where clouds are hit \textit{'face-on'}, similar to the single cloud case.
Both limiting cases can be adjusted in how far the clouds are separated in both the orthogonal and wind-aligned directions (shown in \autoref{res:aligned-no-spacing}, \autoref{res:aligned-with-spacing} and \autoref{res:orthogonal}).
In between the limiting cases, we use a general ellipsoidal (including spherical) configuration of clouds.
Ellipsoids are divided into two subsets classified by the size of their cross section with respect to the wind-axis, depending on whether the semi-major or semi-minor axis is aligned perpendicularly to the wind (\autoref{res:ellipsoidal} and \autoref{res:ellispoidal-shapes}).
Furthermore, we probe an intermediary state of spherical arrangements of the individual clouds in \autoref{res:sphere-non-random} and \autoref{res:spheres-random}.
In the following section \autoref{sec:discussion}, we discuss a new, unifying criterion, that applies to all the morphologies we probed.
Furthermore, in Appendix~\ref{app:periodic} we discuss the results stemming from periodic boxes.

\subsection{A Wind-Aligned Morphology of Clouds} \label{res:aligned}
\subsubsection{Without Inter-Cloud Separation} \label{res:aligned-no-spacing}
\begin{figure*}
\centering
\begin{subfigure}{0.49\textwidth}
        \centering
    \includegraphics[height = .3\textheight]{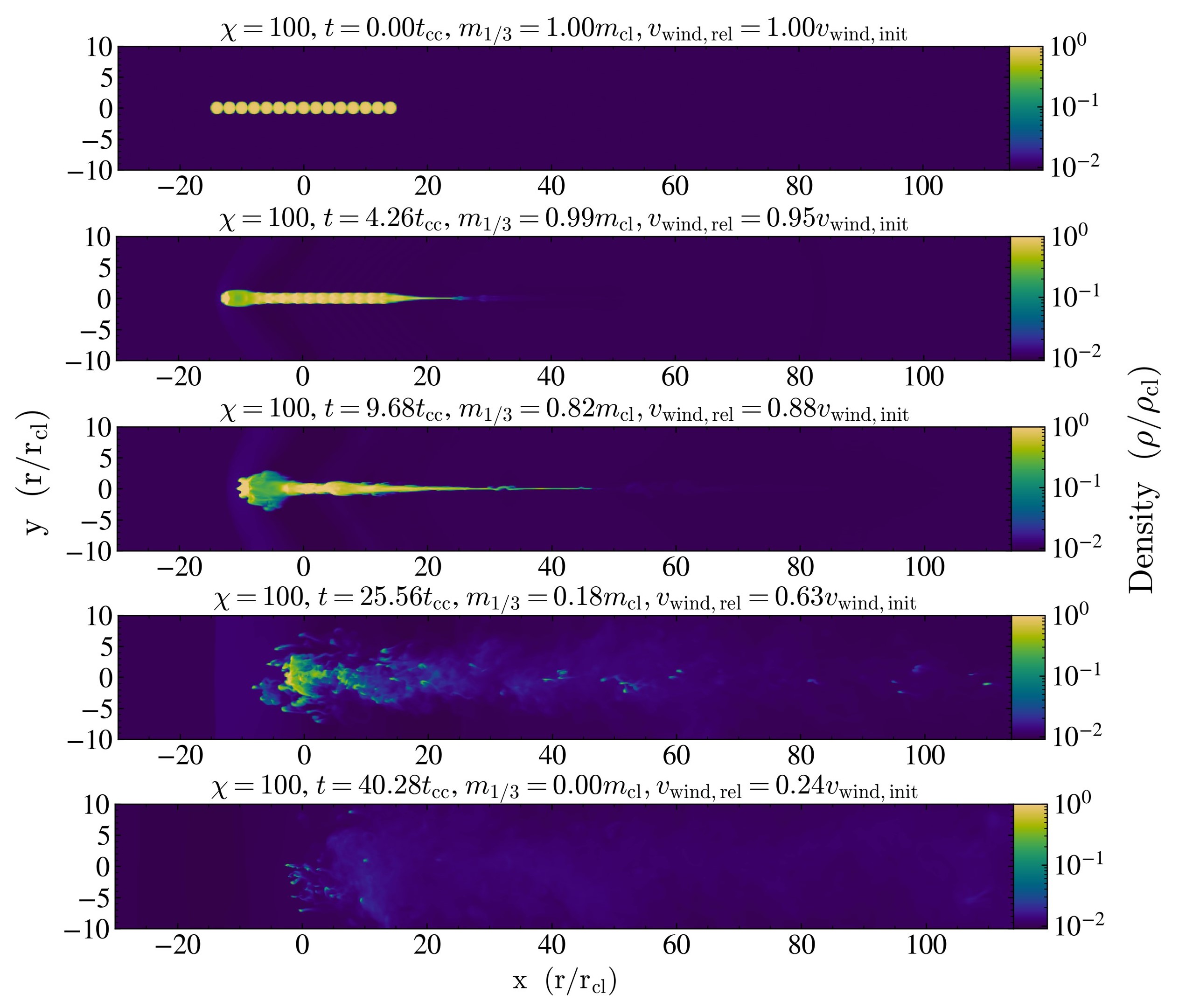}
\end{subfigure}
\centering
\begin{subfigure}{0.49\textwidth}
        \centering
    \includegraphics[height = .30\textheight]{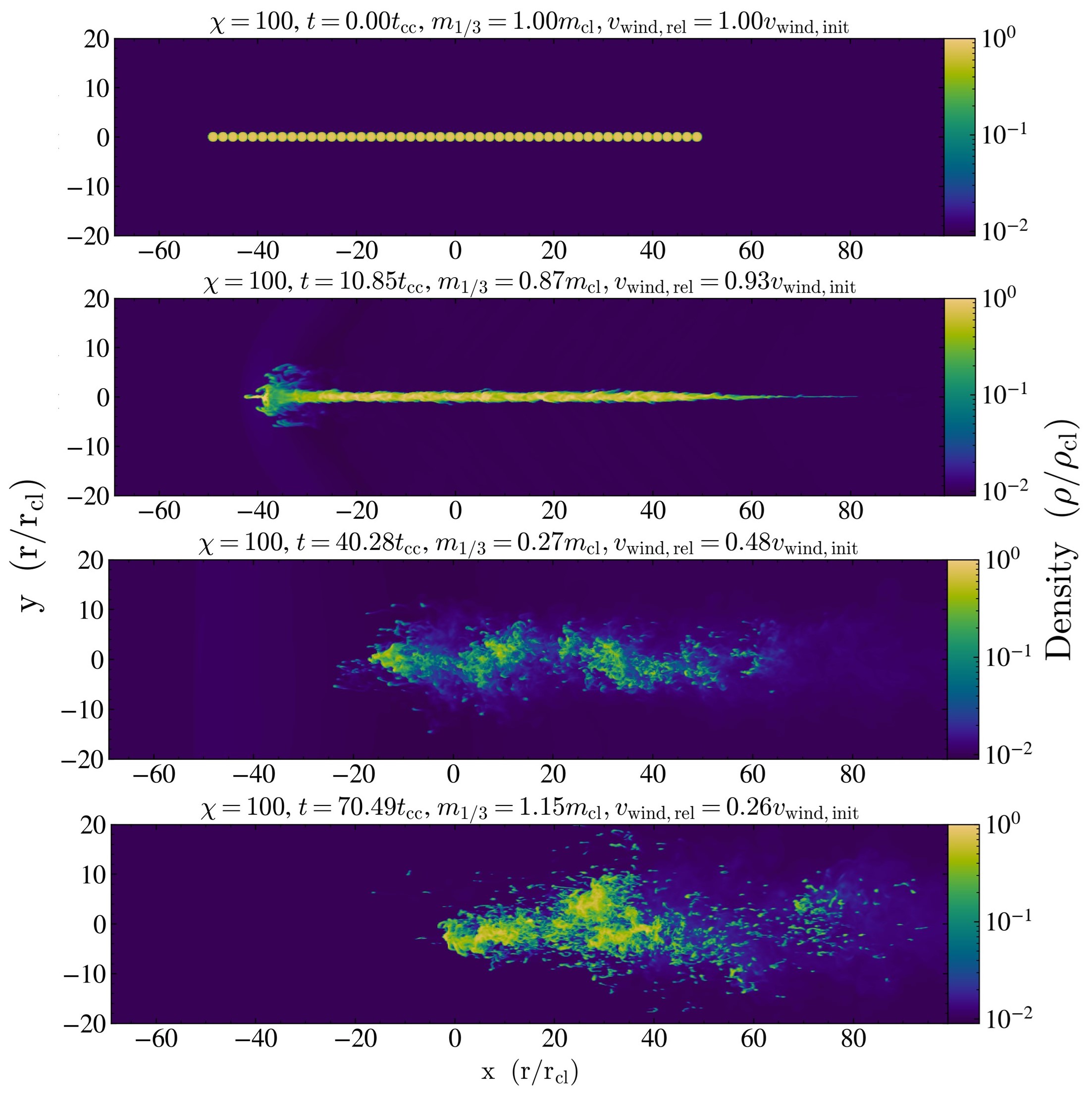}
\end{subfigure}

    \caption{Mass-weighted mean of the density in a non-periodic wind-aligned cloud setup.
There is no spatial separation between the clouds and we visualize only gas mass that is denser than one percent of the initial cloud's density.
On the left side, we see a setup of 15 clouds that are being destroyed while the right-hand side shows early stages of cloud survival and entrainment for a total of 50 clouds. Both initial morphologies form a tail that can enable cloud growth only if enough clouds are placed initially.}
    \label{fig:snapshots_series1_destsurv}
\end{figure*}
\autoref{fig:snapshots_series1_destsurv} shows mass-weighted mean of the density field from two simulations where clouds are arranged directly next to one another; i.e., no inter-cloud distance.
Note, we include mass above a threshold of $1/3 \rho_{\rm cloud}$ towards the \textit{cloud mass}.
The left panel shows virtually no signs of cold gas in the final snapshot, whereas in the right panel, there is a turning point and the cloud continues to cool down more gas by condensation in the final snapshot at around $70 \tcc$.
The left panel shows what we call \textit{destruction} while the right panel leads to \textit{survival}. 
In some other simulations we need to make a different distinction we refer to as \textit{borderline} survival, where the cold gas content decreases but does not fully vanish after some tens of cloud crushing times.

\begin{figure}
    \centering
    \includegraphics[width=\columnwidth]{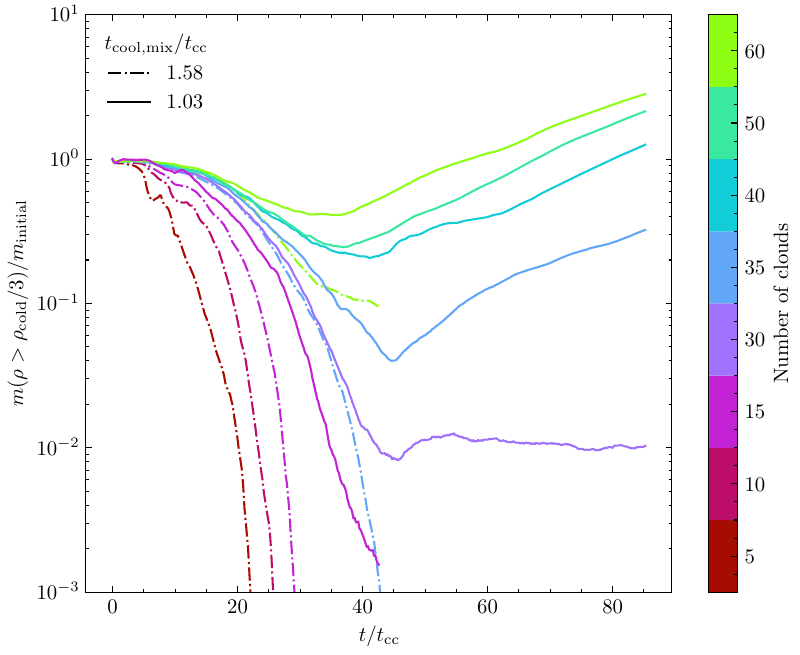}
    \caption{Varying numbers of clouds in a setup that is aligned parallel to the wind direction. The figure shows the relative growth in cold gas mass up to 80 cloud crushing times for different numbers of clouds and different ratios of $\protect t_{\mathrm{cool,mix}}/t_{\mathrm{cc}}$.
The ratio applies only to the single clouds which are placed in a horizontal setup, so that only one cloud faces the wind directly.
In these runs, there is no spatial separation of the individual clouds.
For an increase in total clouds one can see an increasingly well-entrained system of cold gas, the turning point for a fixed ratio of $1.03$ lies somewhere in the region of 30 to 35 clouds.
For a larger ratio, this number increases so that in the range of 35 to 60 clouds a turning point might be found.}
    \label{fig:series1_mass}
\end{figure}

The approach to bring a system, that is eventually destroyed, to survival, involves adding clouds downstream of the wind, until we see an increase in cold gas after some time.
We show the mass evolution diagram, as introduced in \autoref{subsec:morphology}, in \autoref{fig:series1_mass}.
Color-coded are the number of clouds placed in each simulation box, and the line style indicates the value of $\ratioinline$.
We see that a smaller number of clouds leads to certain destruction of the system in both initial configurations.
By increasing the number of clouds, we expect the system to move closer to a state, where entrainment becomes efficient again, and the cold gas mass starts to increase after a while.
The simulations with a lower value of $\ratioinline$ in general need fewer clouds in total to move towards survival.
A threshold of around 35 clouds appears to be sufficient for the lower ratio simulations to survive.
For the higher ratio of the individual cloud, we need at least sixty clouds in a line, to get close to survival, while 30 clouds in a line for the lower ratio, shows an outcome that we call borderline survival.
In contrast to the destruction cases, the cold gas does not vanish as clearly, yet in the long run, we do not find notable growth of cold gas mass. 
The underlying effect driving the borderline evolution might be a mixture of physical and numerical effects.
This work does not feature a detailed analysis of these effects, but a further discussion of numerical difficulties is featured in Appendix~\ref{app:periodic} for periodic boxes.

The general behavior can be explained when revisiting \autoref{fig:snapshots_series1_destsurv}. 
When the wind first hits the leftmost cloud, a tail starts to form, while the rightmost clouds are shielded from the wind.
The cloud mass that interacts with the wind first is pushed towards this tail and starts to mix with it.
If there are enough clouds to form a tail, then the combined cold gas mass is sufficiently large, such that cooling of hot gas via condensation enables entrainment.
This is similar to a single but larger cloud that has some of its mass pushed into the tail by the wind, which then starts to consume hot gas off of it.
Such an argument comes very close to the effective radius argument (see \autoref{subsubsec:effective_radius}), that tells us how much combined gas mass is needed to surpass the ratio threshold of unity and move into the survival regime.
The difference between higher and lower ratios is also clear.
The turnover point for simulations with a lower ratio that show survival is reached after comparable time. 
The number of clouds needed, however, is much larger, as the individual cloud is deeper in the destruction regime (governed by the ratio); therefore, a longer tail, thus more clouds, is needed to surpass the critical gas mass in the tail for entrainment to kick in.

\subsubsection{With Inter-Cloud Separation} \label{res:aligned-with-spacing}
\begin{figure}
    \centering
    \includegraphics[width=\columnwidth]{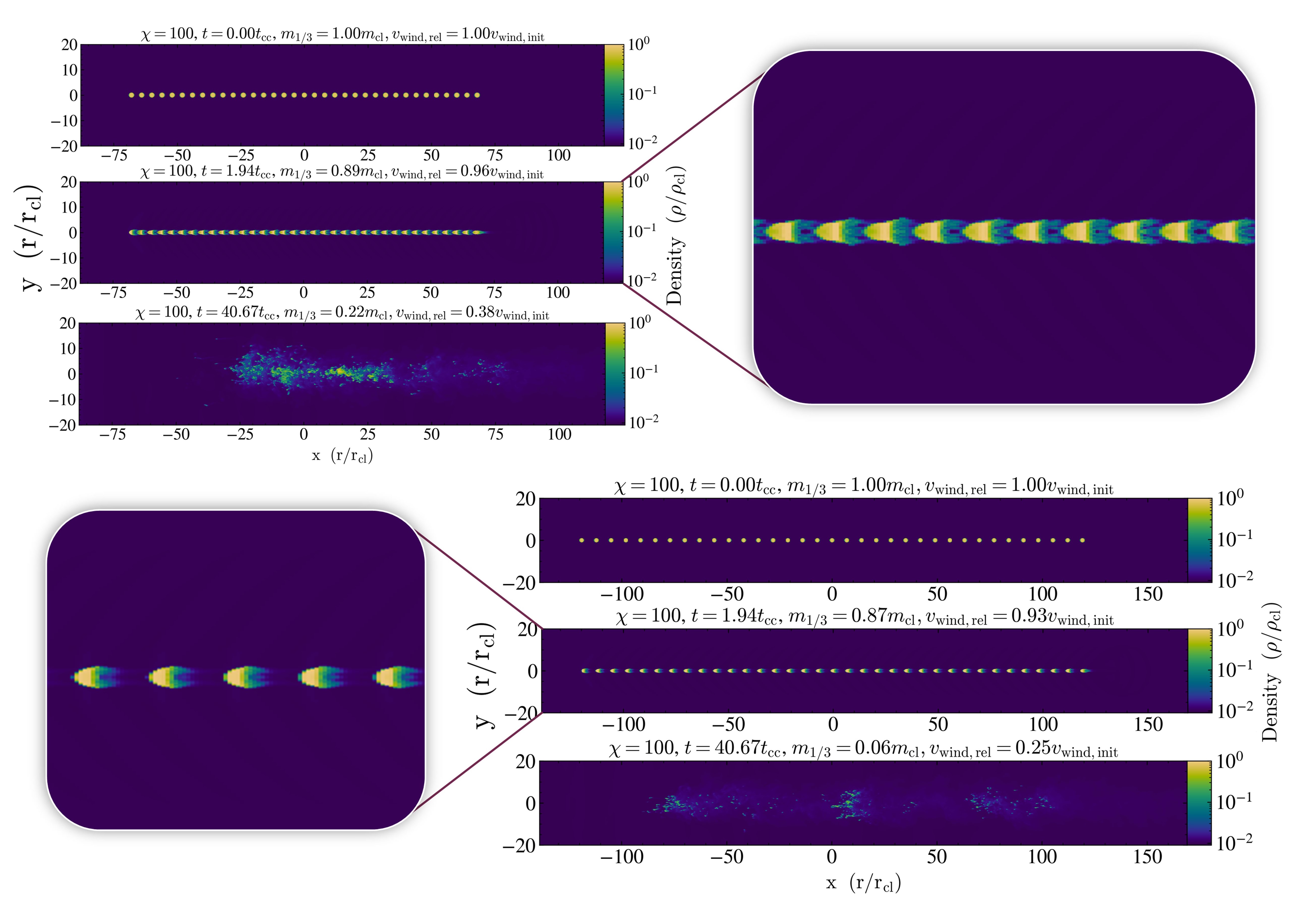}
    \caption{The wind-aligned setup with varying distances between the individual clouds.
The upper panel shows a total of 35 clouds with a spatial separation of $2\rcloud$, while the lower panel shows 35 clouds with a separation of $5\rcloud$.
A crucial point at around $1.94\tcc$ shows why the system in the upper panel survives and the lower one gets destroyed.
The zoom-in reveals how each cloud evolves individually at first, and starts forming a tail.
If closer together, those tails start to mix and combine into a larger contiguous volume of cold gas, that allows for more efficient mass accretion and gas condensation.
}
    \label{fig:snapshots_series1_distance}
\end{figure}

\begin{figure}
    \centering
    \includegraphics[width=\columnwidth]{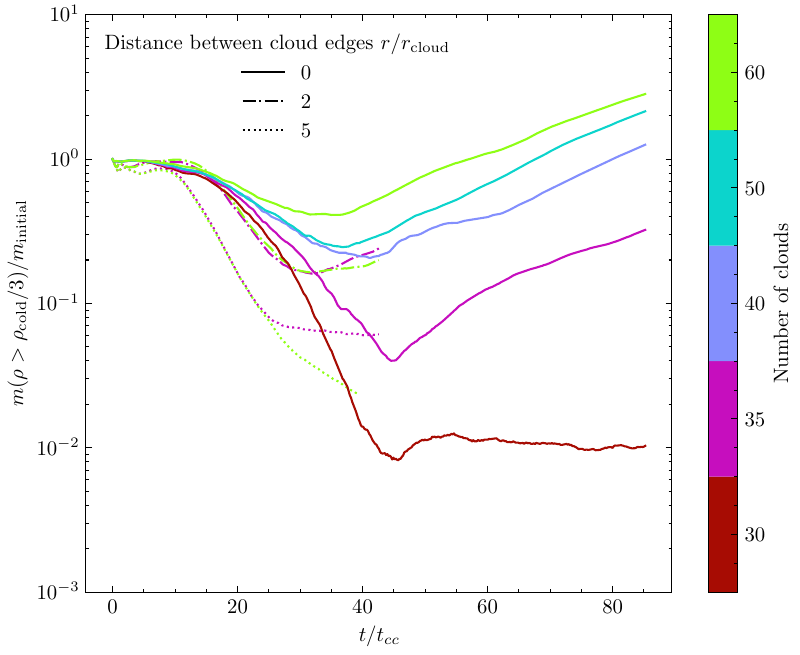}
    \caption{A wind-aligned setup with different numbers of clouds with varying edge-to-edge distances between them (the inter-cloud distance). Shown is the mass evolution of cold gas for up to 80 cloud crushing times.
    Similar to figure \ref{fig:series1_mass} we now introduce a spatial displacement of the individual clouds.
The single cloud parameters stay the same, with a ratio of $\protect t_{\mathrm{cool,mix}}/t_{\mathrm{cc}}\approx 1.03$.
The relative growth in mass is now dependent on both the total amount of cold gas mass and the inter-cloud distance.
Larger inter-cloud distances cause delayed entrainment, which might not be fulfilled at all for 35 and 60 clouds.} 
    \label{fig:series1_dist_mass}
\end{figure}

\autoref{fig:series1_dist_mass} shows the mass evolution of systems with a wind-aligned morphology where the clouds are farther apart.
We measure the inter-cloud distance from edge-to-edge between the individual clouds, and the inter-cloud distance is uniform among each neighboring pair;$1\,\rcloud$ inter-cloud distance corresponds to $3\,\rcloud$ distance from center to center. 

\autoref{fig:series1_dist_mass} shows systems with varying number of clouds and increasing inter-cloud distance from edge-to-edge.
Here, we include the zero distance systems (already shown in \autoref{fig:series1_mass}) with a cloud number of more than 30 for a better comparison.
The zero distance case showed that both a system of 35 and 60 clouds survives, therefore, we take these initial conditions and increase the distance between clouds to 2 and 5 $\rcloud$, respectively.
For the case of $2\rcloud$ distance we find that the turnover point is reached after the same time and the same mass loss.
Similarly, a turnover point is reached for the system of $5\rcloud$ distance after the same time.
The systems with smaller spatial separation start to increase their gas mass for both numbers of clouds while the systems with larger separation continue to decrease in gas mass.
This suggests that increasing the inter-cloud distance decouples the number of clouds from survival and destruction, which only depend on the distance between clouds.
At $5\rcloud$ the outcome is indifferent to the number of clouds, as these evolve individually and their tails do not affect the downstream clouds as they did for the systems in \autoref{res:aligned-no-spacing}.
Similarly, for a distance of $2\rcloud$ the distance remains sufficiently small such that the tails mix into a combined gas volume and accrete mass.
Simulations for $5 \rcloud$ separation show that the 60 cloud configuration loses more mass than its progenitor with fewer clouds.
This is most probably a numerical artifact, and we declare the 35 cloud simulation to be borderline survival, whereas the 60 cloud case is destruction; the main result is that neither of them survive.

\autoref{fig:snapshots_series1_distance} shows the difference between low and high separation.
In both panels we show 35 clouds that are placed with a distance of $2\,\rcloud$ between them in the upper panel and a separation of $5\,\rcloud$ in the lower panel.
The upper panel and zoom-in give evidence that the tail of each cloud affects the downstream clouds by stirring up its environment, possibly enhancing the mixing and cooling ability of the composite gas.
At a certain separation, as seen in the lower panel, the tail cannot influence the neighboring cloud and the single cloud's destruction dictates the system's evolution.

\subsubsection{Summary of the Wind-Aligned Setup} \label{sec:summary-wind-aligned}

\begin{figure}
    \centering
    \includegraphics[width=\columnwidth]{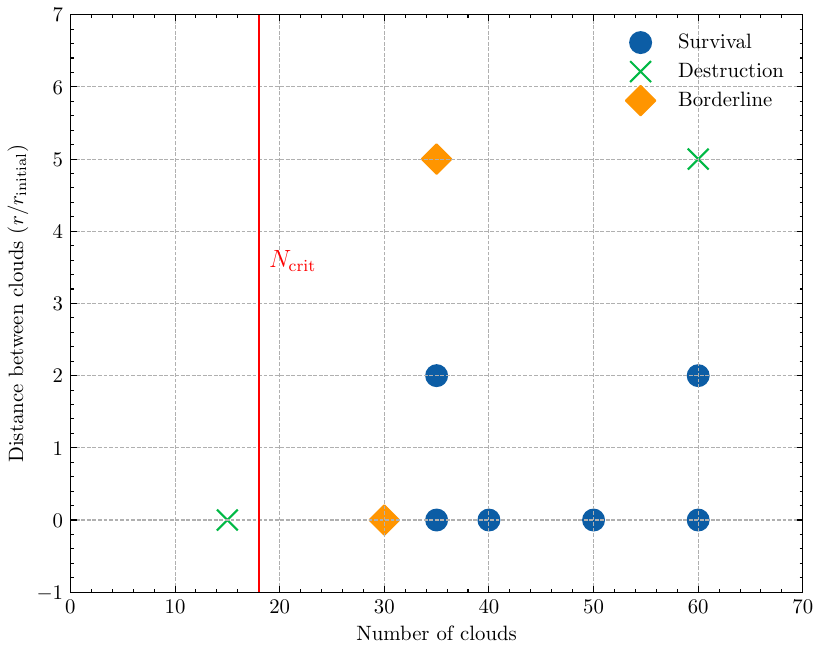}
    \caption{Overview of surviving and destroyed systems in a wind-aligned setup with and without inter-cloud separation.
    Shown are the number of clouds on the horizontal axis and the pairwise distance between neighboring clouds on the vertical axis.
    We find survival and destruction based on the outcome of the simulations, by investigating the mass evolution diagrams.
    Imprinted in red is the critical threshold in number of clouds as dictated by the effective radius and the related mass limit criterion (see \ref{subsubsec:effective_radius}).
    For zero separation, this criterion is almost fulfilled, but with increasing distance between the clouds, the evolution of the system decouples from the number of clouds and the criterion fails to predict the correct outcome.}
    \label{fig:series1_survdiagram}
\end{figure}

In \autoref{fig:series1_survdiagram}, we show a type of diagram, we call a \textit{survival diagram}, to investigate the previously implied mass limit that decides survival versus destruction.
The horizontal axis displays the number of clouds placed in the initial system's configuration, and the vertical axis shows the inter-cloud distance.
We use marker types to encode the simulation results.
With a ratio $\ratioinline=1.03$, we initially proposed a mass limit criterion, where one needs to place 18 individual clouds in the system to surpass the mass needed for survival \autoref{subsubsec:effective_radius}.
The red vertical line in \autoref{fig:series1_survdiagram} shows this boundary between expected survival and destruction predicted by such a mass limit criterion.
For systems without inter-cloud separation, this criterion seems to be somewhat fulfilled; yet, as described earlier, the evolutionary outcome decouples from the number of clouds with increasing distance between them.
For the systems with $5\rcloud$ separation, survival is not seen even far above the supposed threshold.
In general, the mass-limit criterion is not fulfilled, even for this most simple setup.
A new criterion must take into account the physical shape of the configuration of clouds, in addition to the amount of gas within a composite volume. 
  
\subsection{An Orthogonal Morphology of Clouds} \label{res:orthogonal}

\begin{figure*}
    \centering
    \includegraphics[width=\textwidth]{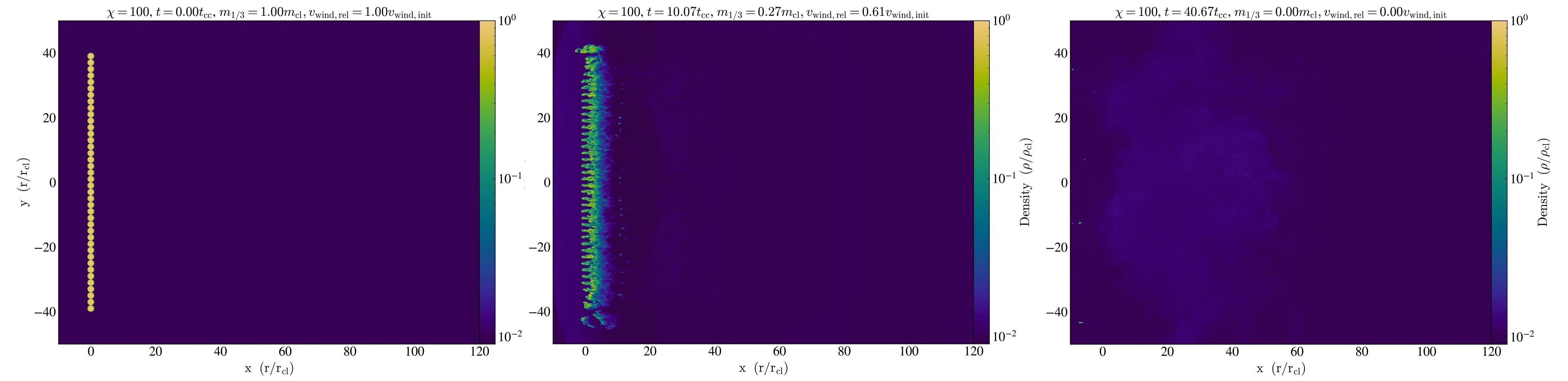}
    \caption{The initial setup and time evolution for a non-periodic setup of 40 clouds stacked orthogonal to the wind direction.
We see a quick destruction of the initial clouds. In the right panel in the lower left corner, we see small artifacts of cell-sized, high density regions of numerical nature.}
    \label{fig:series2_nonPdest}
\end{figure*}
\begin{figure}
    \centering
    \includegraphics[width=\columnwidth]{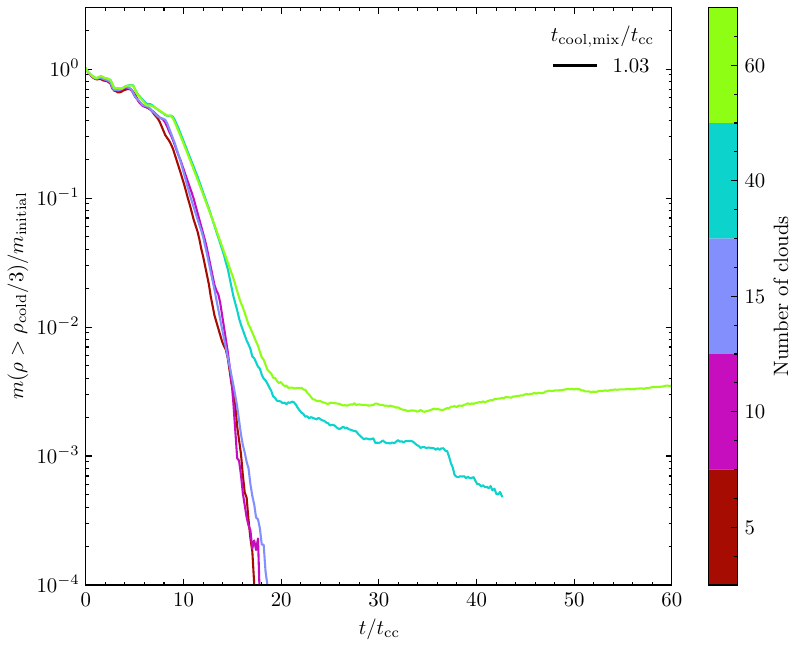}
    \caption{The relative cold mass evolution up to 60 cloud crushing times for systems where the clouds are placed in a row orthogonal to the wind without inter-cloud separation.
The single cloud ratio is $\ratioinline\approx 1.03$.
For an increase in the total number of clouds, we see the turnover point separating destruction and survival lies between 40 and 60 clouds, yet even the 60 cloud setup does not show proper mass growth.
All systems are eventually destroyed in the wind, but starting at 40 clouds, the mass loss is slowed, hinting that more clouds might support cloud growth.
Each individual cloud shows a destruction similar to the isolated, single-cloud crushing simulation.}
    \label{fig:series2_nonP_mass}
\end{figure}

In \autoref{fig:series2_nonP_mass}, we show the evolution of cold gas mass for systems with clouds placed next to one another orthogonal to the wind direction.
On the horizontal axis, we show the time evolution until $60 \tcc$, and the vertical axis shows the amount of cold gas normalized to the initial mass.
The colors depict an increasing number of clouds, and there is no inter-cloud distance between neighboring clouds (now in the direction orthogonal to the wind).
Until 15 clouds in a row, destruction is very clear. 
Starting at 40 clouds, we see slower mass loss and stagnation at a level below 1\% of the initial cold gas mass; we attribute borderline survival to these systems.
The fluctuations at a very low mass are caused by numerical artifacts that remain in the box, preventing the cold gas mass from vanishing entirely.

When taking a closer look at the mean density in \autoref{fig:series2_nonPdest}, where we show a system of 40 clouds in a single row without inter-cloud separation, we see small and dense droplets in the lower corner of the right panel at the size of a few cells.
These artifacts are of numerical nature; regardless, the overall system shows borderline survival after $40\tcc$.
Similarly, for the 60 cloud system, we find borderline survival.

Given that \autoref{fig:series2_nonP_mass} shows a slightly slower mass loss for systems with a high number of clouds, we expect some mechanism exists that can drive cloud growth even for the orthogonal setup.
The clouds' tails start mixing in the composite volume downstream of the wind and might affect clouds that are already within the composite volume, just like in the wind-aligned setup in \autoref{res:aligned}.
Additional mass mixed in from above and below might support cloud growth.

\begin{figure}
    \centering
    \includegraphics[width=\columnwidth]{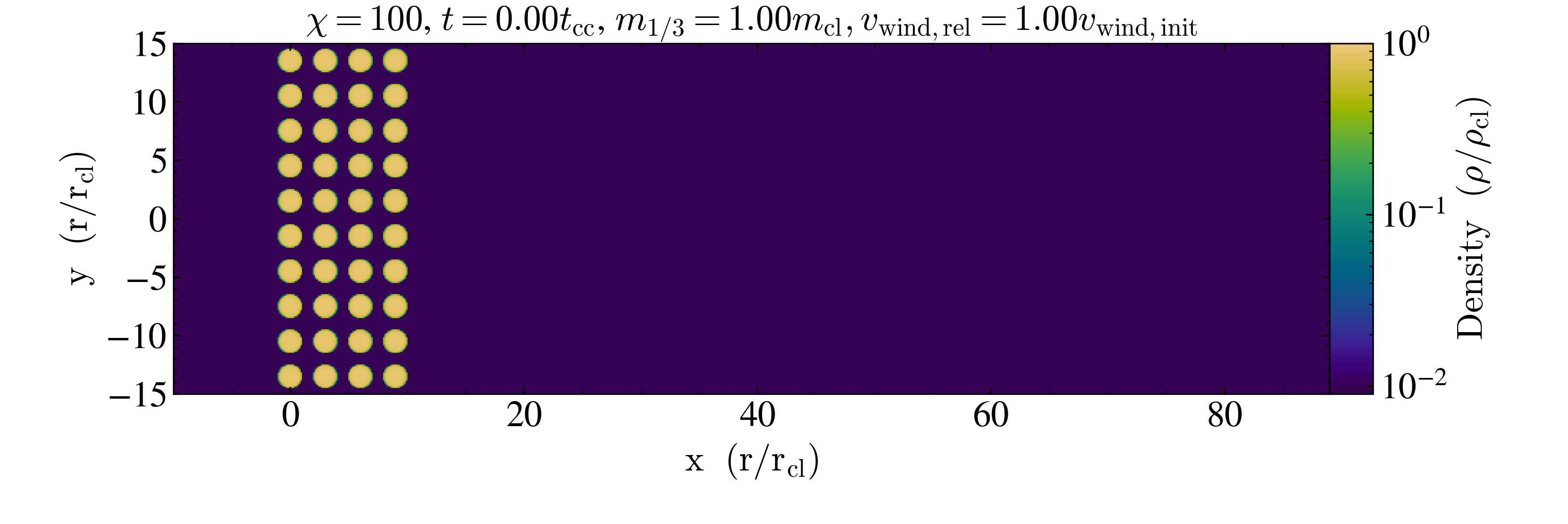}
    \caption{Initial condition for a multi-cloud setup orthogonal to the wind shown as a mass-weighted mean density.
In positive and negative y-directions we introduced a periodic boundary and we steadily increase the number of rows placed behind one another to vary the number of clouds, that are not directly exposed to the wind.
In this case, we set up 4 rows with a spatial separation of $1\,\rcloud$, resulting in an infinitely wide bar of clouds.}
    \label{fig:series2_yP_inital}
\end{figure}

In \autoref{fig:series2_yP_inital} we show the initial configuration of a system that combines the wind-aligned and orthogonal setups.
Note, the y-direction of the simulation box is now periodic in both directions.
By placing additional rows of clouds in the wind's path, we feed mass in the composite volume, which is initially shielded from the wind and enriches the mixed gas, supporting cloud growth.
Now, once again, the clouds have a spatial separation, to make them comparable to results from \autoref{res:aligned-with-spacing}.

\begin{figure}
    \centering
    \includegraphics[width=\columnwidth]{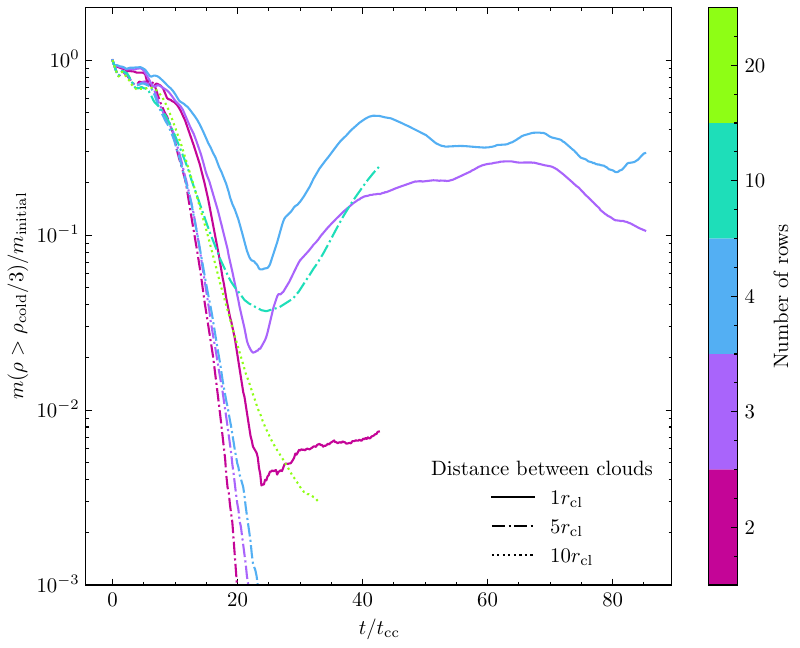}
    \caption{Cold gas mass evolution for a system with periodic boundaries in the positive and negative y-direction, creating a bar-like structure. The line styles encode varying inter-cloud separation between neighboring clouds in both x- and y-direction, while the color imprints the number of rows placed downwind. For the smallest spatial separation, a total of 3 rows is sufficient for survival, while we need at least 10 rows for a larger distance of $5\rcloud$. For even larger distances, not even a total of 20 rows is sufficient to see survival of the system. The solid lines show the mass decreasing at around 50 and 70 $\tcc$, as a result of cold gas leaving the simulation box and pressure decreasing in the system (likely an error of numerical nature) respectively. Compared to \autoref{res:aligned-with-spacing} we need far fewer clouds downwind to see survival because of additional mass fed into the stirred up volume by neighboring clouds.}
    \label{fig:series2_yP_mass}
\end{figure}

\autoref{fig:series2_yP_mass} shows the mass evolution of the dense and cold gas varying the number of rows and inter-cloud distances (with a periodic setup).
The dashed lines show systems with an inter-cloud separation of $1\rcloud$.
We find that in a y-periodic box more than 3 rows lead to notable cloud growth, while two rows are not sufficient.
Going back to the wind-aligned setup, where we need at least 35 clouds in a row for the survival of a system, the mass feeding in an orthogonal and infinitely wide system supports the mass accretion up to a point, where 3 rows of clouds are sufficient.
The mass loss at around $50 \tcc$ in the 4 rows system is caused by some of the cold gas leaving the simulation box upstream, and the decline in both 3 and 4 row systems at around $70\tcc$ is the result of a pressure drop in the entire simulation box, making cooling less efficient.
In addition, at late times, in a co-moving system (with the wind), where little gas is stirred up and creating turbulent mixing layers, cooling is far less prominent.

With an inter-cloud separation of $5\rcloud$, dash-dotted lines in \autoref{fig:series2_yP_mass} show that there is a clear turnaround in cold gas mass for 10 rows.
All systems with fewer rows or even larger separations -- 20 rows are not sufficient for a system of $10\rcloud$ cloud separation to survive -- are destroyed quickly.

The $5\rcloud$ case with 10 rows shows again how mass feeding into the composite volume brings a system to survival.
Comparing it to the wind-aligned configuration where 60 clouds in a line with the same inter-cloud separation in (compare \autoref{fig:series1_dist_mass}) could not support survival gives evidence for the mass feeding argument.
We feature several more periodic boxes in \autoref{app:periodic}, these however come with more numerical difficulties.

\subsection{Generalized Ellipsoidal Morphology} \label{res:ellipsoidal}

Both configurations introduced in the previous sections are idealized simplifications that will not be found in nature.
We want to move one step closer towards a realistic density distribution found in astrophysical systems.
Therefore, the following sections will introduce a new terminology.
In order to assimilate a realistic probability density function (PDF) we have to find discretizations of any global morphology in terms of smaller clouds.
We often refer to the global morphology as the \textit{enclosing volume}, arranged by a varying number of smaller clouds (more clouds usually result in an arrangement more similar to the global morphology).
This allows us to make use of the survival criterion of the individual clouds $\ratioinline<1$ and effectively probe different \textit{volume filling fractions} of various global morphologies.

\begin{figure}
\centering
\begin{subfigure}{0.34\columnwidth}
        \centering
    \includegraphics[width=\columnwidth]{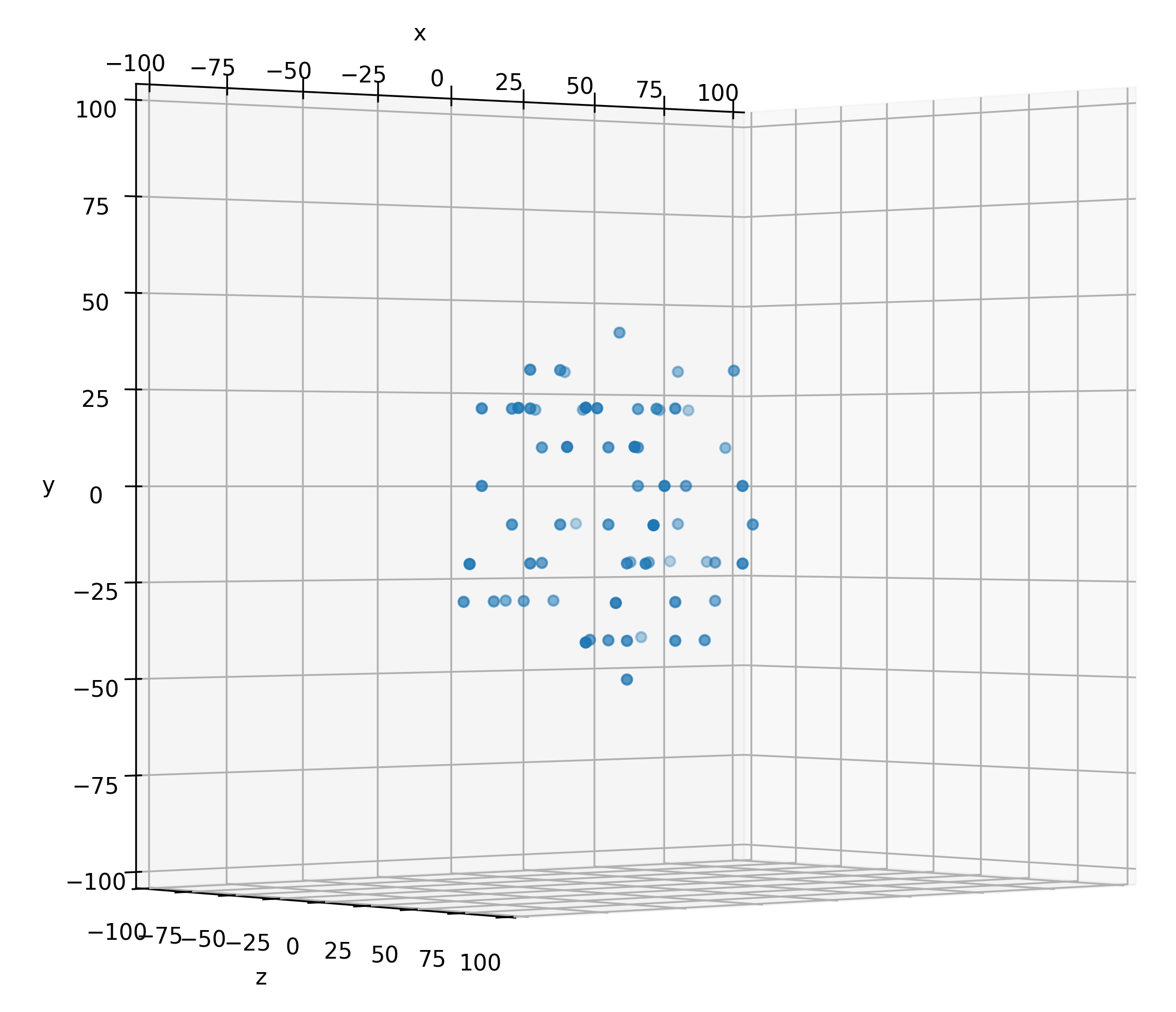}
\end{subfigure}
\begin{subfigure}{0.64\columnwidth}
        \centering
    \includegraphics[width=\columnwidth]{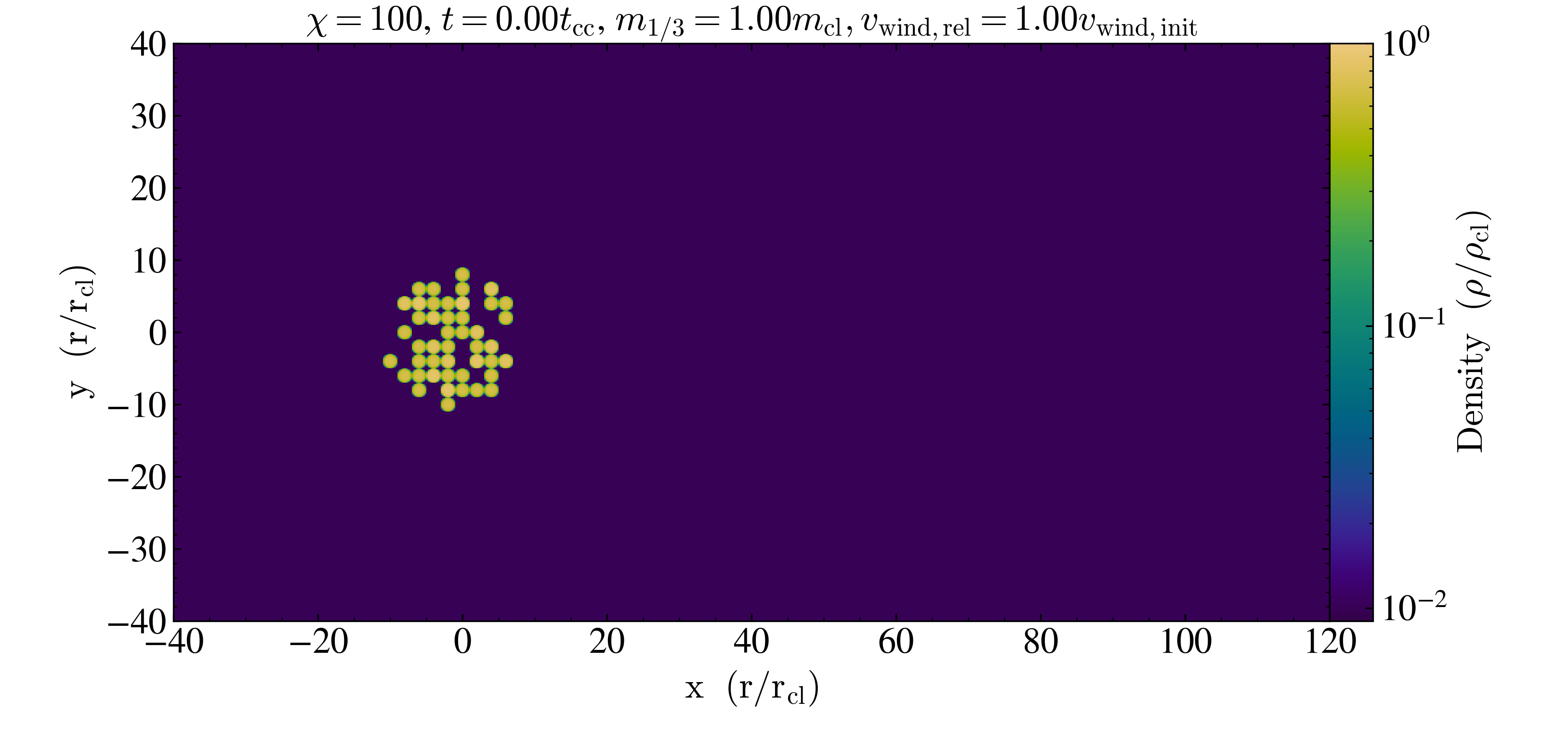}
\end{subfigure}
    \caption{Initial conditions and setup of the individual spheres for a general ellipsoidal (in this case spherical) enclosing volume.
The left panel shows a 3D render of the cloud configuration (clouds not to scale).
The right panel shows the same configuration as a mass-weighted mean density, showing the grid-like nature of our sampling of clouds.
The wind is flowing along the x axis of the simulation box and the ratio remains at $\ratioinline\approx1.03$ for now.
This system has a configuration of 64 clouds placed in an enclosing sphere of radius $10\rcloud$.
}
    \label{fig:series3_initalcond}
\end{figure}

\autoref{fig:series3_initalcond} shows a 3D-visualization of the initial conditions, where we construct a spherical global morphology only with smaller clouds.
For the larger part of this section, we construct a grid in the enclosing volume with $1\rcloud$ spacing and sample clouds from all grid nodes.
Sampling more clouds will increase the volume filling fraction $f=V_\mathrm{gas}/V_\mathrm{enc} = N_\mathrm{cloud}V_\mathrm{cloud}/(4/3 \,\pi a_x a_y a_z)$, where $a_x, a_y$ and $a_z$ are axes of the enclosing ellipsoid and $V_\mathrm{cloud}=4/3\pi r_\mathrm{cloud}^3$.
However, a caveat of this approach is some regularity in the sampled configuration of clouds, which we address in \autoref{res:spheres-random}.

Whereas the mass limit criterion failed to predict the survival of a system (see \autoref{sec:summary-wind-aligned}), we investigate a possible threshold in volume filling fraction that can predict destruction versus survival.
Importantly, the volume filling fraction remains indifferent to the global morphology.
However, one may wonder what happens, if we use many small clouds that add to the same total mass of a few larger clouds, yet make up the same enclosing shape.
Therefore, we are effectively investigating if survival depends on fragmentation of the global morphology.

\subsubsection{Spheres of Different Filling Fraction} \label{res:sphere-non-random}

Starting with a configuration similar to what we show in \autoref{fig:series3_initalcond} we vary both the size of the enclosing sphere (an ellipsoid with $a_x=a_y=a_z$) and the number of clouds we put into the enclosing volume.
The individual clouds still satisfy $\ratioinline\approx 1.03$ and thus sampling more clouds into the same enclosing volume will increase the volume filling fraction accordingly. 

\begin{figure}
    \centering
    \includegraphics[width=\columnwidth]{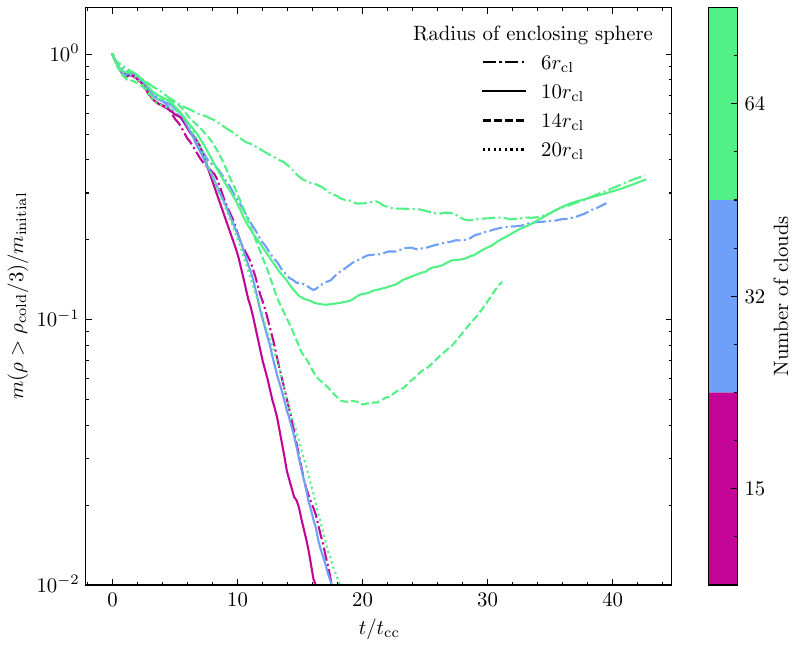}
    \caption{Gas mass evolution for the general ellipsoidal setup with spherical enclosing volumes.
    The line style indicates the radius of the enclosing sphere and the number of clouds is encoded in the color.
    As all individual clouds across the systems have the same mass and ratio $\ratioinline\approx1.03$, we see that more clouds, thus gas mass, generally lead to survival, if packed tight enough.
    For too few clouds, we find that even a smaller enclosing volume is not sufficient, but once enough mass is in the system initially, a more compact packing leads to survival of the system.}
    \label{fig:series3_spheres_mass}
\end{figure}

\autoref{fig:series3_spheres_mass} shows the cold gas mass evolution for various systems.
The color encodes the number of clouds put into the enclosing volume, which we differentiate by the line styles, ranging from 6 to 20 $\rcloud$ radius of the enclosing sphere. 
For 15 clouds, both 6 and 10 $\rcloud$ radii of the bounding sphere were not sufficient to see the systems' survival.
The next step, 32 clouds, surpasses our previously implied mass limit criterion, and we see that the smallest enclosing sphere leads to a tight packing of clouds, with a large volume filling fraction, and supports survival.
As before, the mass limit criterion only works for very compact systems with low inter-cloud separation (naturally fulfilled by a system with high volume filling fraction).
Systems with 64 total clouds can be seen to move from certain destruction to clear survival when decreasing the radius of the  circumscribing sphere.
While $20 \rcloud$ enclosing radius still leads to destruction, we find that with tighter packing, the turnover point shifts towards later times, and the maximum mass loss becomes lower (from 4\% via 10\% to 25\% of the initial mass being leftover at the lowest point).

\begin{figure}
    \centering
    \includegraphics[width=\columnwidth]{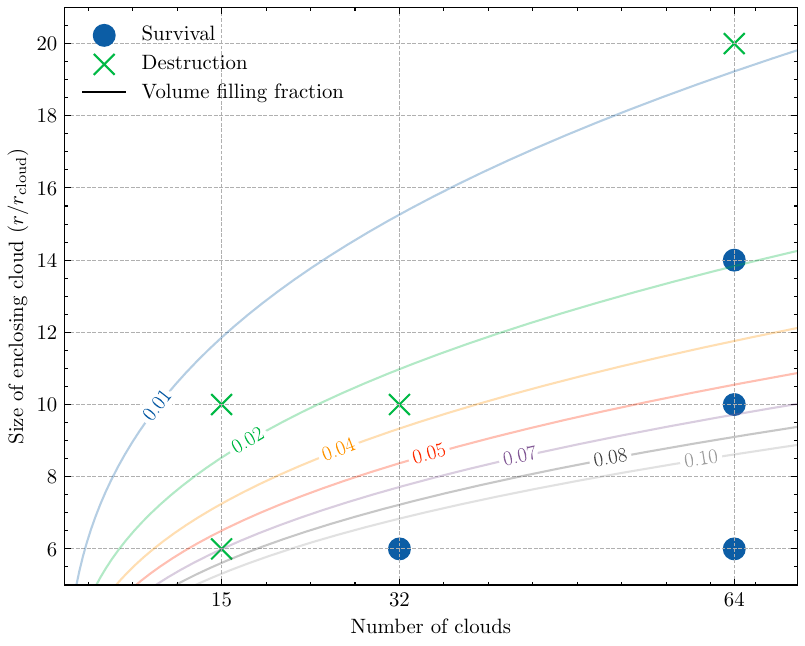}
    \caption{The survival diagram for a general spherical setup.
The lines drawn into the figure show contours of equal volume filling fractions.
The markers show different evolutions for the simulated setups.
The diagrams spans the parameter space for the number of clouds placed into an enclosing spherical volume of a given radius.
We separate destruction from survival and see that it cannot be an equal volume filling fraction that causes survival or destruction.
Most remarkable are runs that lie on the same contour for volume filling fractions, but show different evolutions.
Increasing the fragmentation by inserting more clouds but keeping the volume filling fraction equal, suggests a general trend towards survival.}
    \label{fig:series3_equilines}
\end{figure}

In \autoref{fig:series3_equilines} we show a survival diagram for the above described spherical configuration.
The horizontal axis shows the number of individual clouds placed in the enclosing sphere, whose radius is shown on the vertical axis.
We do not see any borderline behavior anymore.
In addition, to check a possible volume filling fraction threshold for survival, we show contours of equal volume filling fraction in the survival diagram.
While it appears there is a distinct region in parameter space where systems survive, this area does not coincide with a cutoff in volume filling fraction.
As an illustrative example, we compare the system with 64 clouds in a $14 \rcloud$ radius enclosing sphere with the one with 32 clouds and a $10\rcloud$ enclosing sphere.
Both show virtually the same volume filling fraction however the system with more clouds survives.

\subsubsection{Ellipsoids of Different Shape} \label{res:ellispoidal-shapes}

\begin{figure}
\centering
\begin{subfigure}{\columnwidth}
        \centering
    \includegraphics[width=\columnwidth]{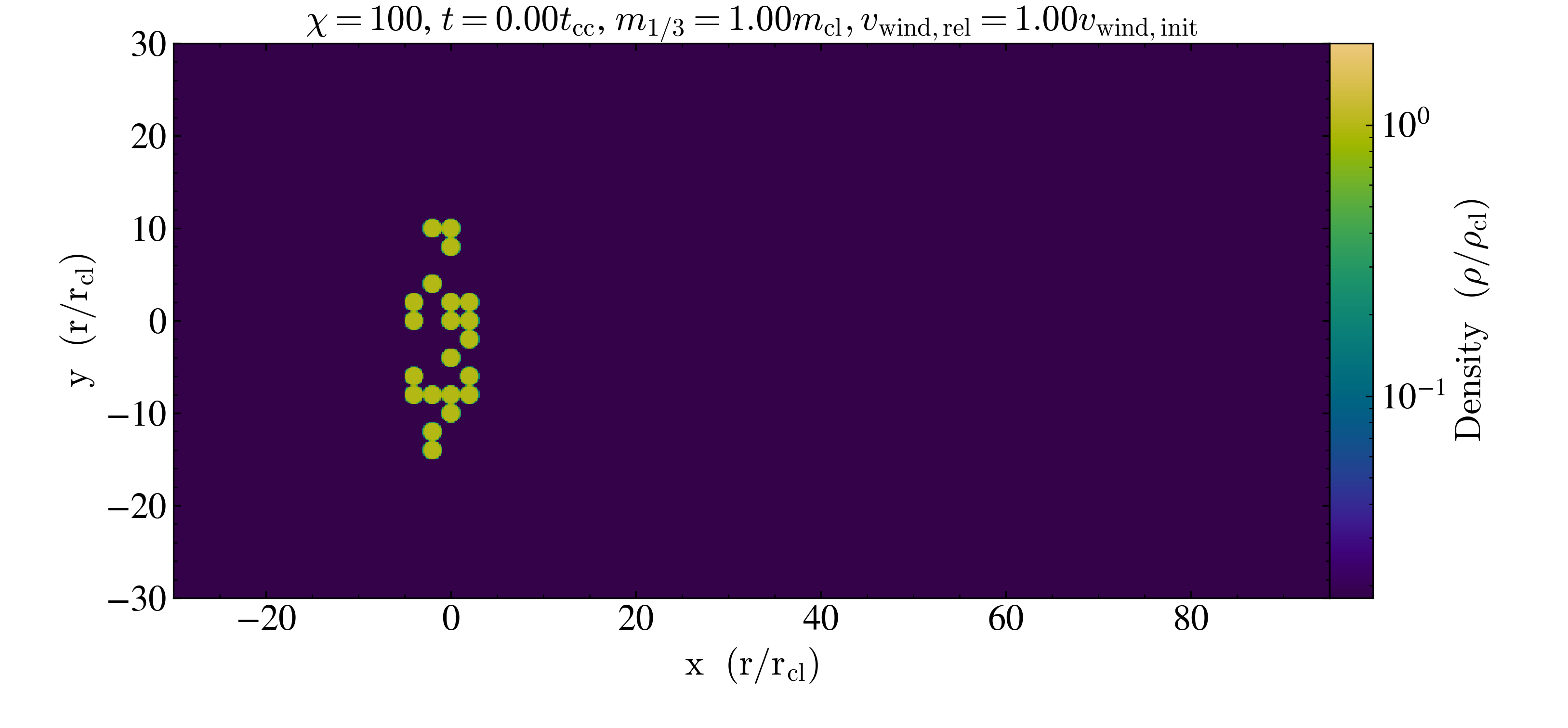}
\end{subfigure}
\centering
\begin{subfigure}{\columnwidth}
        \centering
    \includegraphics[width=\columnwidth]{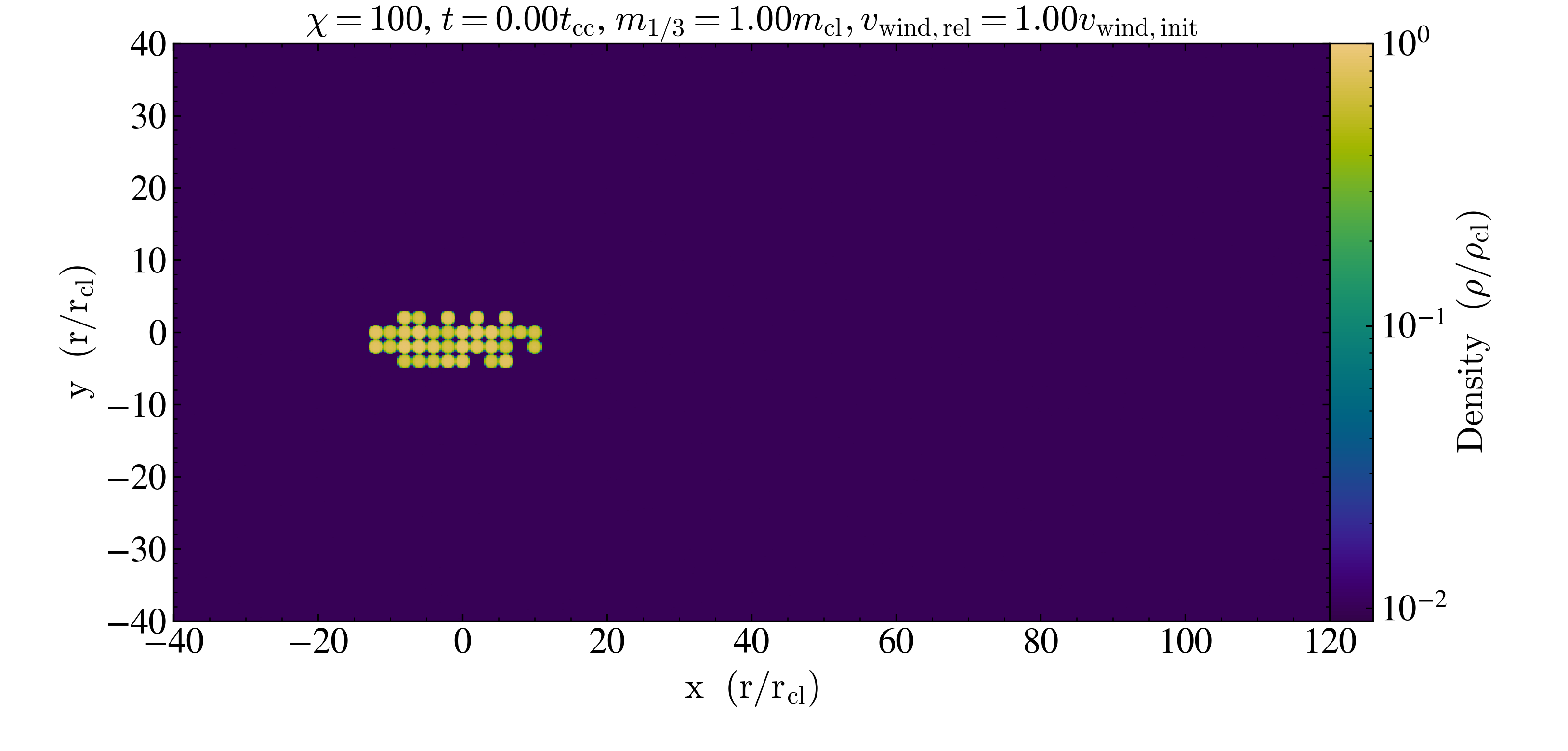}
\end{subfigure}
        \caption{Initial configuration of systems with an ellipsoidal global morphology. 
        We show the mass-weighted mean density of two systems with 64 clouds in total.
        Due to sampling clouds from a grid, all clouds overlap perfectly in this type of projection, causing a loss of spatial information in the z direction.
        In general we differentiate ellipsoids of high(er) wind area (upper panel; $a_y>a_x=a_z$) and low wind area (lower panel; $a_x>a_y=a_z$).}
    \label{fig:series3_ell_initial}
\end{figure}

In \autoref{fig:series3_ell_initial}, we show the initial conditions for two ellipsoids in mass-weighted mean density.
However, note that the mean density fails to visualize the three-dimensional nature.
Matching our fiducial case, the individual clouds still fulfill $\ratioinline\approx 1.03$.
In general, we differentiate ellipsoids of high(er) wind area (upper panel; $a_y>a_x=a_z$) and low wind area (lower panel; $a_x>a_y=a_z$).
The upper panel in \autoref{fig:series3_ell_initial} is more similar to the orthogonal configuration while the lower panel assimilates the wind-aligned setup.

\begin{figure}
    \centering
    \includegraphics[width=\columnwidth]{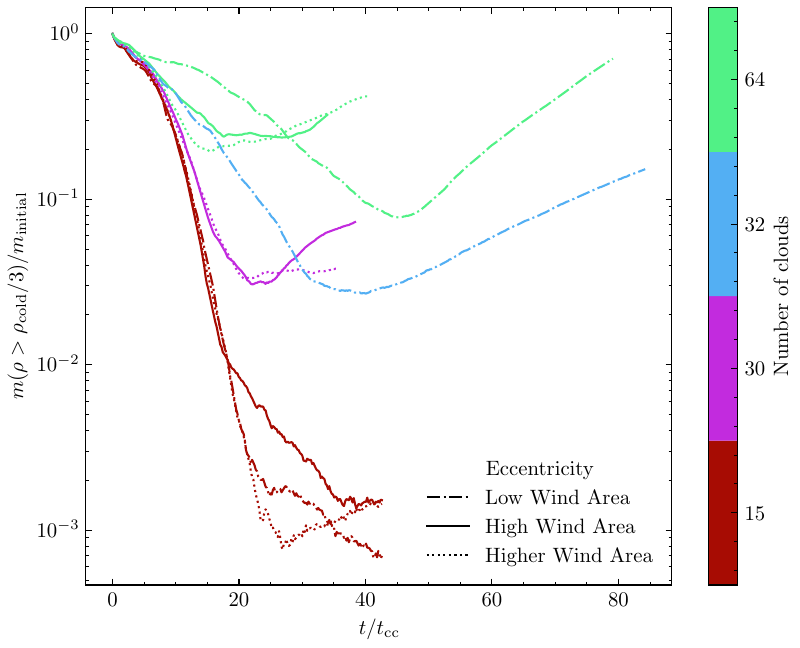}
    \caption{Mass growth diagram for ellipsoidal setups of various kinds.
We show the amount of cold gas mass normalized to the initial total mass.
Color-coded are different numbers of clouds that are used to create the enclosing shape of the global morphology.
The individual clouds reside at $\ratioinline\approx1.03$ and would be destroyed on their own.
High and low wind areas only differentiate between ellipsoids that face the wind with their larger or smaller cross-section (see \autoref{fig:series3_ell_initial}).
The general trends are an earlier turnaround point for high wind area runs, as these initially encounter the wind more strongly and start to grow via condensation much earlier.
Similar to a wind-aligned system, the low wind area sees a delayed turnaround point at a generally lower total gas mass, which nevertheless grows in mass afterwards.
The fate of systems appears to be governed only by the number of clouds again, instead of the shape of global morphology.
}\label{fig:series3_ellips_mass}
\end{figure}

\autoref{fig:series3_ellips_mass} shows the cold gas mass evolution for a set of ellipsoids with varying alignment with the wind and numbers of clouds.
For 15 clouds, as seen previously in this work, and thus below the mass limit threshold of $\approx 18$, we see definite destruction for all alignments.
At a threshold of 30 clouds, both alignments for high(er) and low wind area already show survival. 
Interestingly, they all drop to about 3\% of their initial cold gas mass at their lowest.
For the low wind area setup, the larger semi-axis is then parallel to the wind direction. 
Consequently, the mass loss is delayed compared to the high(er) wind area systems.
We see similar behavior in terms of turnover point and delayed destruction for systems of 64 total clouds too.
The low wind area system shows slower mass loss initially, reaches the minimum gas mass at much later time than the high(er) wind area systems, and drops to about 8\% of the initial system's mass compared to 20\% in the high wind area setup.

The general evolution of these systems is explained by the first impact the wind makes on the individual clouds.
The high(er) wind area setups are similar to orthogonally placed systems of clouds, causing a much more rapid destruction at early times of the original configuration.
This comes at a loss of cold gas mass at first, generates a heavily stirred up environment directly after, and uses the cooling ability of turbulent mixing layers to the fullest, enabling rapid cloud growth only shortly after the first impact.
On the other side, the low wind area systems are similar to a wind-aligned setup where the cloud gets dragged along with the wind lazily, and loses cold gas slowly.
Again, the wind-aligned setups show a very solid cold gas growth at a later point, powered by consuming hot wind gas through condensation.
The systems we show in \autoref{fig:series3_ellips_mass} indicate that survival once again decouples from the actual shape and only depends on the number of clouds, which fits well with the volume filling fraction ansatz for a new survival criterion.
\subsubsection{Fragmentation in a Randomized Spherical Morphology} \label{res:spheres-random}
\begin{figure}
    \centering
    \includegraphics[width=\columnwidth]{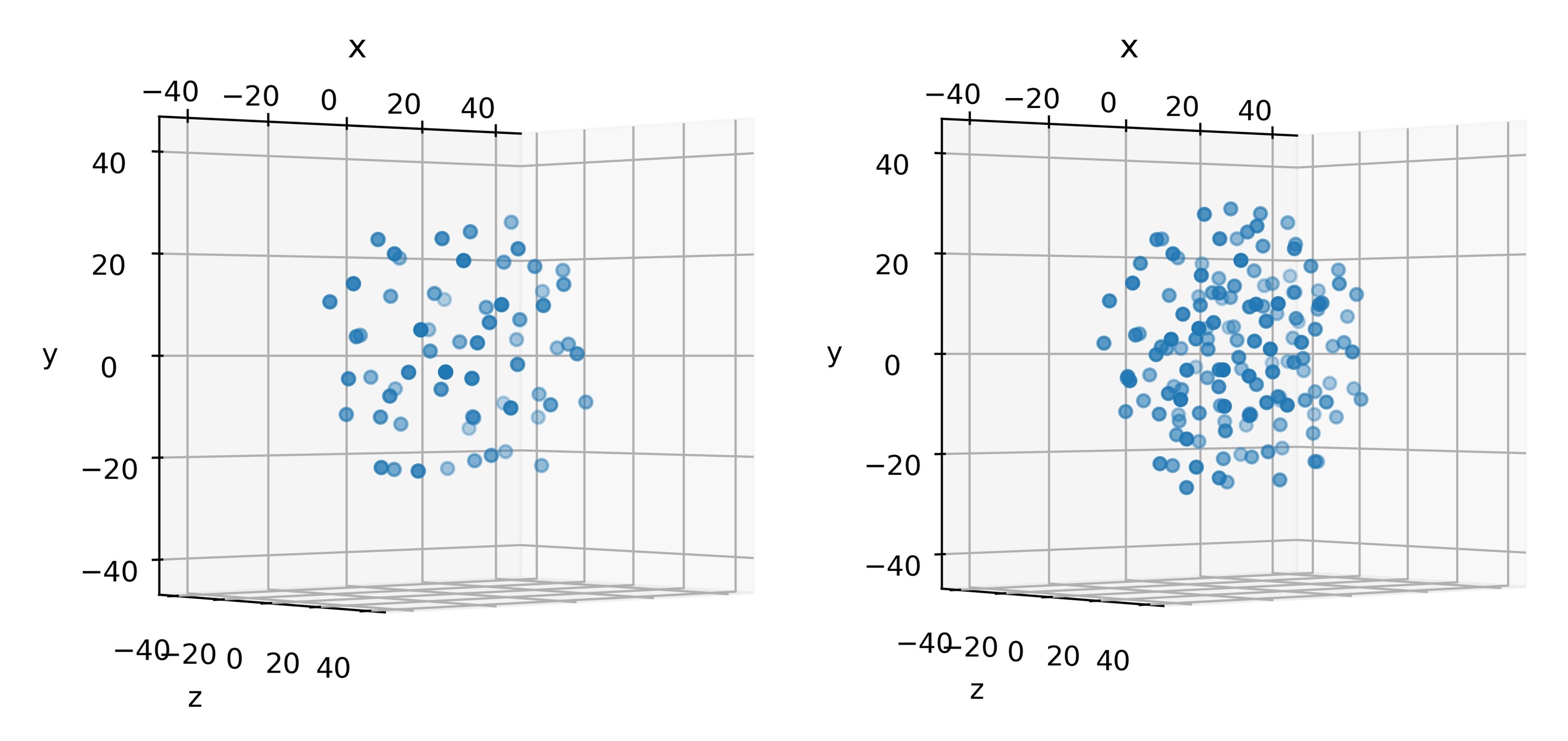}
    \caption{The cloud setup for a more random distribution.
Both initializations show a volume filling fraction of 0.15, while the left-hand picture uses 64 clouds and the right-hand figure uses 150 clouds to make up that filling fraction.
The size of the enclosing sphere stays the same for both scenarios.
The sizes of the dots do not represent the actual size but merely show the position of the cloud's center.}
    \label{fig:series4_intial}
\end{figure}
\autoref{fig:series4_intial} shows the approach we use to implement increasing degrees of fragmentation in purely spherical setups, while employing a more random sampling technique.
Both panels show the position of individual clouds' centers placed in an enclosing sphere of radius $30\rcloud$, and the dots are not to scale with the actual clouds in the simulations' boxes.
The systems we show have the same volume filling fraction of 15\% but the right panel uses many smaller clouds to fill the volume, whereas the left panel uses fewer, but larger clouds.
Since scaling the clouds changes the ratio $\ratioinline$, we choose a few different values for the ratio.

\begin{figure}
    \centering
    \includegraphics[width=\columnwidth]{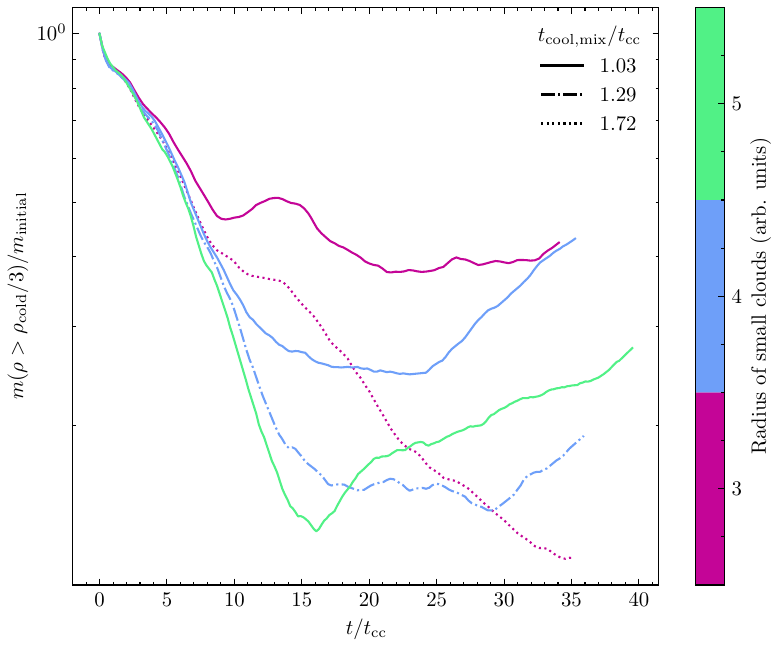}
    \caption{Cold gas mass evolution for a random and fragmented spherical setup.
All the simulations are set up to have the same volume filling fraction 15\% but with varying sizes of the small clouds (in arbitrary units) and therefore varying fragmentation.
The line style denotes $\ratioinline$. This effectively increases the physical size of the system, but comes at a larger degree of fragmentation at a fixed volume filling fraction.
A general trend is that with increasingly fragmented systems, the mass loss is smaller until the turnaround point is reached for all systems after roughly the same time.
Systems where the ratio is at a different value, also show a turnover point but behave differently in general.}
    \label{fig:series4_massevolution}
\end{figure}

In \autoref{fig:series4_massevolution} we show the cold gas mass evolution for a few systems.
Color encodes the radius of the individual clouds where smaller size corresponds to more clouds needing to be placed in the enclosing sphere to get to the same volume filling fraction of 15\% for all systems.
When rescaling the cloud sizes, the ratio of cooling time to the cloud crushing time changes.

Changing $\ratioinline$ effectively acts like we place more clouds -- of the same size as before -- into a much larger enclosing volume.
While the physical size and mass of such a system would increase greatly, we keep the volume filling fraction constant, which creates a system at a much higher degree of fragmentation.

We investigate the systems with a ratio of 1.03, as in all systems before, and find that with increasing fragmentation (smaller cloud radii) the system loses less mass before reaching a turnover point.
One driver of this evolution might be, that many clouds at $\ratioinline\approx 1.03$ as in the purple dashed line, come at a much greater total physical cloud mass, greatly surpassing the mass limit criterion of 18 clouds \autoref{subsubsec:effective_radius}, and thus see great mass growth.

In the extreme case of only a few clouds at a volume filling fraction 15\% (e.g., a wind orthogonal setup of two clouds only) we can say that the system is likely destroyed, so pushing the initial condition towards higher fragmentation will lead to survival.
We leave a convergence study to show that this effect is of physical nature to future work.

\section{Discussion} \label{sec:discussion}
\subsection{A multi-cloud survival criterion} \label{sec:survival_criterion}

In previous sections of this work we made some initial naive estimates for a critical threshold of volume filling fraction or an effective radius.
The idea of an effective radius and equivalent mass limit introduced in section \ref{subsec:morphology} turned out to be unsuitable as a unified survival criterion (see, e.g. \autoref{fig:series1_survdiagram}).
Likewise, we found a simple volume filling fraction is unable to capture details of multi-cloud dynamics, whereas the fragmentation and shape of the cloud configurations appears to be more important (see, e.g. \autoref{fig:series3_equilines} or \autoref{fig:series4_massevolution} with a constant volume filling fraction but different evolutions). 

Our approach to finding a multi-cloud survival criterion is based on an \textit{effective volume filling fraction $F_V$} that scales with various problem specific parameters, like $\chi$, $\rcloud$, $\Mwind$ and $\tcoolmix$, that define cloud survival in the single cloud case.
The basic requirements for a multi-cloud survival criterion are the following
\begin{enumerate}
    \item The single cloud survival criterion $\ratioinline<1$ must be captured.
    \item It must adapt to the environment (e.g.  $\chi$, $\rcloud$, $\Mwind$, $\tcoolmix$).
    \item The spatial configuration (aligned vs. orthogonal vs. ellipsoidal) must be taken into account.
    \item It needs to differentiate between a tightly/maximally packed and loosely packed configuration (i.e. different inter-cloud distances in all configurations).
    \item Stochasticity in configurations like the ellipsoidal ones may be among the cases where the criterion breaks down.
\end{enumerate}

\begin{figure}
    \centering
    \includegraphics[width=\linewidth]{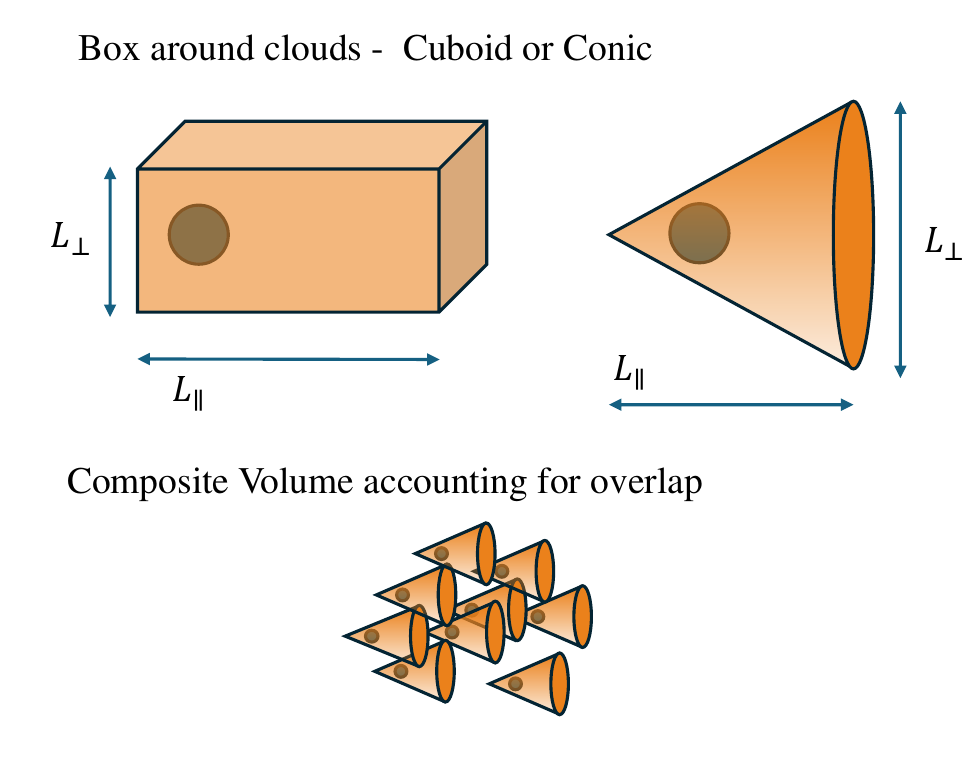}
    \caption{A schematic overview of the process of creating a composite volume out of enclosing boxes. The upper panel shows different shapes of the enclosing box volume and how length parameters are encoded within. The lower panel shows an array of clouds and possible overlap between neighboring boxes which encodes spatial information depending on each configuration via the overlap.}
    \label{fig:comp-overview}
\end{figure}

To encode the spatial configuration in the criterion we employ a post-processing routine that takes the initial position of clouds and constructs a \textit{box} around them.
Then we construct a composite volume already incorporating the overlap between boxes -- a crucial step, as it alters the entire volume of boxes if clouds are placed closer together (closer packing means smaller composite volume with the same number of clouds).
The effective volume filling fraction then simply consists of the fraction of cold gas volume of the clouds and the total composite volume $F_V = V_\mathrm{cl}/V_\mathrm{comp}$.
The choice of a suitable box shape also affects the resulting effective volume filling fractions.
We use two parameters to control the size of the box $L_\perp$ in the y and z directions (orthogonal to the wind direction) and $L_\parallel$ (parallel to the wind direction) where the individual cloud sits at the leftmost side of the box, centered at $L_\perp/2$ from the edge.
\autoref{fig:comp-overview} provides a visual overview of the above described geometry.

In the following, we will make use of a \textit{conical-shaped} box and we provide results for a \textit{cuboid-shaped} box in Appendix~\ref{app:cuboid_fv}.
The conical shape has the advantageous effect that the spread of the cloud's tail downwind is being covered more accurately.
Cold gas slowly spreads to the sides as it becomes co-moving with the wind.
The initial phase of disruption where the first shocks go through the cloud will spread the gas mass towards the side, but in this phase the mass feeding into the neighboring cloud's tail is not important just yet.
When the systems slowly become co-moving, the mass feeding plays a crucial role and this is where the \textit{conic spread} shows its effects.

The base of the cone has a radius of $L_\perp/2$ and its height is $L_\parallel$ accounting for a total volume per box of $V_\mathrm{cone}=1/3\pi(L_\perp/2)^2 L_\parallel$.

To connect the new criterion to the well-known single cloud criterion we have to choose a proportionality to the single cloud length scales in our new length scales $L_\perp$ and $L_\parallel$.
The simplest assumption is that both scale with the cloud radius $L_\perp\propto\alpha\rcloud$ and $L_\parallel\propto\beta\rcloud$.
This, however, immediately violates the initial requirement (i) as the box scales with the cloud size of a single cloud, and thus every cloud, regardless of $\ratioinline$ results in the same volume filling fraction $F_V$ making it unsuitable.

The scaling we choose is  $L_\perp=\alpha \rcrit$ and $L_\parallel=\beta \rcrit$ with $\rcrit=\tcoolmix\chi^{-1/2}\vwind$ as in the single cloud survival criterion derived from $\ratioinline$.
This fulfills all four requirements (i)-(iv) from above.

The single box's volume is given by $V_\mathrm{box}=1/3\pi(L_\perp/2)^2 L_\parallel = 1/12\pi \alpha^2\beta \rcrit^3$ and we show the limiting cases we use to determine a limiting $F_V$.
\begin{enumerate}
    \item For the single cloud setup with $\rcloud>\rcrit$, thus survival, we find 
    \begin{equation}
        F_V \gtrsim \frac{\rcloud^3}{ \alpha^2\beta \rcrit^3}\stackrel{\ratio=1}{=} \frac{1}{\alpha^2\beta}
    \end{equation}
    in order for the cold gas to survive.
    \item The second requirement is fulfilled by the choice of scaling the enclosing volume with $\rcrit$ (which depends on additional parameters; cf. Eq~\eqref{eq:survivalcriterion}).
    \item Accounting for a potential overlap between neighboring boxes we effectively incorporate spatial configuration. 
    Additionally the choice of a conical box penalizes configurations, where many clouds are placed orthogonally but there is no additional gas mass downwind to feed into the tail.
    At the same time, this choice rewards configurations where there is lots of overlap in the downwind direction and thus a good chance of surviving.
    The reward is in the form of an increased volume filling fraction when more clouds are placed between the already existing ones, increasing the cold gas mass, but the composite volume changes little to none due to a high degree of overlap.
    \item A system that is qualified for survival is a system of $N_\mathrm{cl}$ clouds that are tightly packed (maximum overlap) and the single cloud does not survive $\rcloud\ll\rcrit$.
    The volume filling fraction is
    \begin{equation}
         F_V \sim \frac{N_\mathrm{cl}\rcloud^3}{ \alpha^2\beta \rcrit^3} \stackrel{N^{1/3}\rcloud\sim \rcrit}{=} \frac{1}{\alpha^2\beta} \, .
    \end{equation}
    A configuration of loosely packed clouds with little to no overlap will show a filling fraction that is similar to 
    \begin{equation}
        F_V \sim \frac{N_\mathrm{cl}\rcloud^3}{N_\mathrm{cl} \alpha^2\beta \rcrit^3} \stackrel{\rcloud\ll \rcrit}{\ll} \frac{1}{\alpha^2\beta}
    \end{equation}\
    and reduces to the single cloud case for no overlap.
\end{enumerate}

From the above reasoning, we expect a \textit{critical threshold} of 
\begin{equation}
    F_{V,\mathrm{crit}} \sim \frac{1}{\alpha^2\beta}
\end{equation}
where the numerical pre-factor is solely defined by the shape of the box, i.e., $(4/3\pi)/(1/12\pi)=16$ for a conical and $4/3\pi$ for a cuboid box.

\begin{figure*}
    \centering
    \includegraphics[width=\textwidth]{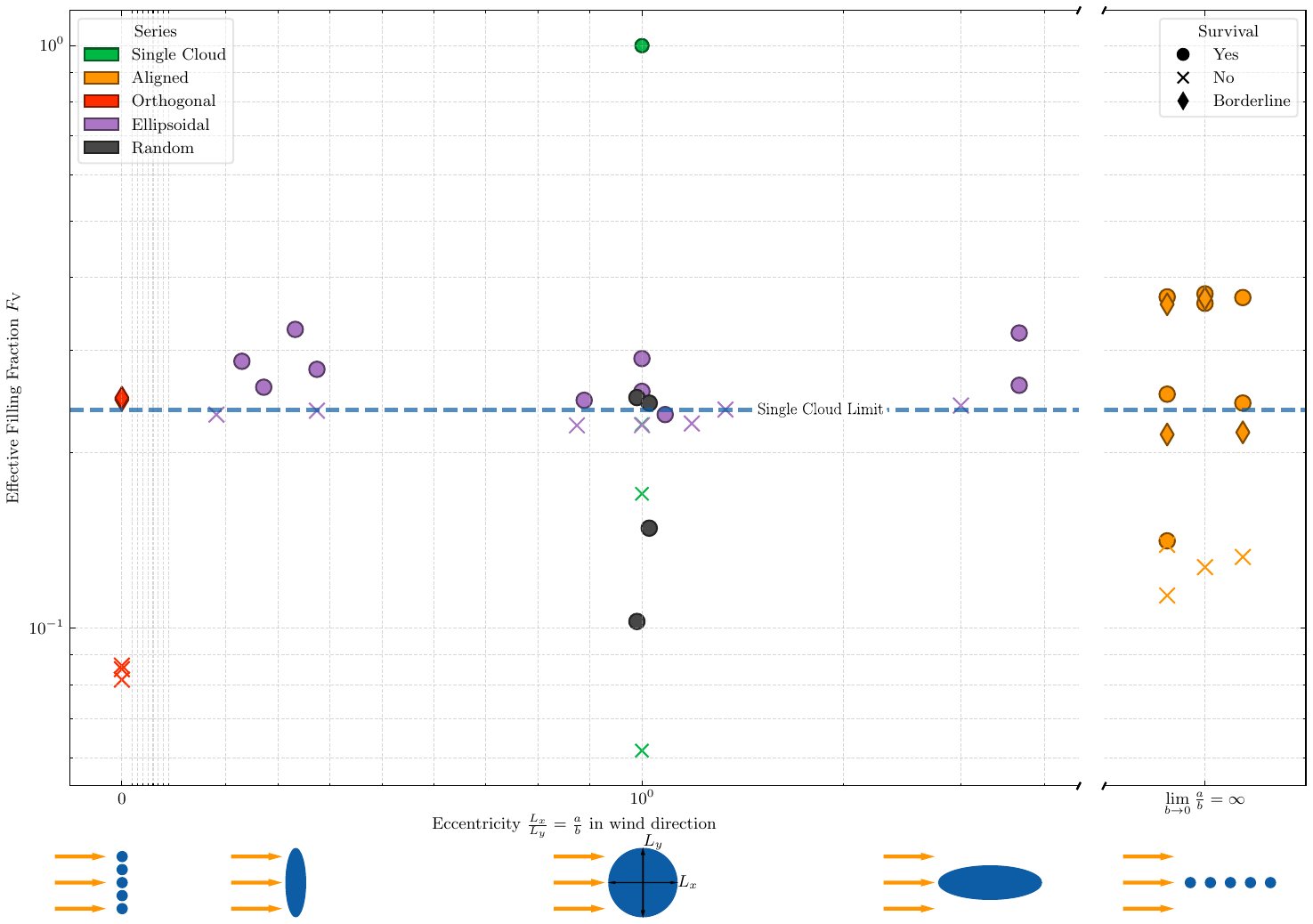}
    \caption{Overview of all morphologies and their effective volume filling fraction. We show the effective volume filling fraction for each configuration based on a conic box (base diameter $L_\perp=\alpha\rcrit=3\rcrit$, height $L_\parallel=\beta\rcrit=7.5\rcrit$) around each individual cloud versus the inferred eccentricity of the configuration. $F_V$ is cut off at a value of 1, where setups of individual clouds with a cloud radius larger than the critical radius (thus $\ratioinline<1$) result in a value much larger than unity. The measure for eccentricity is simply a fraction of the length in the wind-aligned direction and the length in the wind-orthogonal direction, where all the wind-aligned configurations are interpolated to $\infty$ in their limit (additionally there is a random jitter around $a/b=\infty$ to better separate close data points). On the other side, eccentricities of zero are exclusive to the wind-orthogonal setup. The theoretical threshold $F_{V,\mathrm{crit}}=16\alpha^{-2}\beta^{-1}$ implied by the single cloud configuration shows a very clear separation between survival and destruction of a multi-cloud system. The only outliers are the random setup with higher values of $\ratioinline$ and wind-aligned setups with higher $\ratioinline$. We find that the measure $F_V$ can link single-cloud survival theory to multi-cloud morphologies.}
    \label{fig:unified_cone}
\end{figure*}

\autoref{fig:unified_cone} shows the resulting overview of effective volume filling fractions compared to the ellipticity of the configuration for all systems investigated for this work. 
On the horizontal axis we show the eccentricity calculated directly from the initial cloud configuration with $e=L_x/L_y=a/b$ ranging from 0 to $\infty$. 
Values at 0 are the orthogonal setup, while values at $\infty$ are all part of the wind-aligned setup (the actual values $a/b$ are not infinite, but in the limit they approach it). 
A single cloud and likewise an ensemble of clouds in spherical shape take the value 1. 
The cone parameters for the enclosing boxes are $L_\perp=3\rcrit$ and $L_\parallel=7.5\rcrit$ with $F_{V,\mathrm{crit}}\approx 0.24$. 
We chose these values of $L_\perp$, and $L_\parallel$ by visual inspection of similar overview figures but note that the values appear reasonable for the initial extent of the cold gas in a single cloud run.
We find that almost all surviving systems and most of the borderline evolutions fall above the line set by the single cloud criterion $F_{V,\mathrm{crit}}=16\alpha^{-2}\beta^{-1}\approx0.24$ (note that $F_V$ is set to a maximum of 1, as some single cloud systems with a low ratio $\ratioinline$ would fall above a value of 1).
Both random ellipsoidal runs, which have a smaller value of $\ratioinline$,  are outliers.
In those cases, the cloud size is much smaller than $\rcrit$ compared to other systems; therefore they have a smaller cloud volume yet a large composite volume, lowering the effective filling fraction.
One outlier in the wind-aligned setup is a setup with 60 clouds and a higher ratio $\ratioinline\approx1.58$ (see \autoref{fig:series1_mass}).
Such a case may be attributed to stochastic effects, and even then, the difference to the theoretical threshold is only at about $\sim 0.1$.
For an overview where the composite volume has been constructed with \textit{cuboid} boxes, see \autoref{app:cuboid_fv}.

\subsection{Comparison to previous work}
`Cloud crushing' or the `cloud in a windtunnel' setup might be (alongside a single mixing layer) the most basic, and well studied example of multiphase gas dynamics in astrophysics.
While early studies (e.g., \citealp{nittmann1982dynamical,Klein_McKee_Colella_1994}) did not include radiative cooling, and thus studied the rate of destruction of dense gas ablated by a hot wind, this changed with increasing computational power.
Later studies included more physical effects such as cooling as well as increased resolution \citep[e.g.][]{Cooper_Bicknell_Sutherland_Bland-Hawthorn_2009,Scannapieco_Brueggen_2015,Schneider_Robertson_2017}.
Specifically, the inclusion of cooling -- as well as longer simulation boxes and the focus on the tail evolution where mixing occurs and new, cold material condenses -- led to the fact that some studies found surviving cold clouds \citep[e.g.][]{marinacci2010mode,gritton2017condensation,armillotta2017survival} and eventually to the survival criterion Eq.~\eqref{eq:survivalcriterion} \citep{Gronke_2018}
\footnote{In fact, as described in \S~\ref{sec:survival_criterion}, the discussion of what is the correct survival criterion is still ongoing, see, e.g., \citet{kanjilal2021growth,Farber_Gronke_2022, Abruzzo_Fielding_Bryan_2023}.}.

Importantly, these studies focused on the survival of individual clouds (albeit sometimes with complex morphology, see, e.g., \citealp{Schneider_Robertson_2017,Gronke_Oh_2020}). 
However, naturally, the ISM of galaxies is more complex and already early studies \citep[e.g.][]{Cowie_McKee_Ostriker_1981} discussed the importance of several clouds being hit by a supernovae shock due to `shielding' and other synergetic effects. Later studies such as \citet{Poludnenko_Frank_Blackman_2002,Aluzas_Pittard_Hartquist_Falle_Langton_2012,Forbes_Lin_2019} studied these effects numerically and find, overall, that they can prolong the lifetime of the cold mass. However, due to the non-inclusion of cooling or the focus on the `head', (i.e., a relatively short simulation domain and / or duration), these studies did not find mass growth. \citet{Forbes_Lin_2019} find, for instance, that using an infinite number of clouds in the wind direction with no inter-clump spacing can prolong the lifetime of the cold gas versus the single-cloud case by a factor of $\sim 3$. However, because they do not include radiative cooling, the cold gas mass eventually decreases to zero.\footnote{Interestingly, \citet{Forbes_Lin_2019} find a critical inter-cloud spacing (they measure from center to center) of $\delta/r_{\rm cl}\approx 16$ above which they find `rapid destruction'. Between their $\delta\approx8$ and $\delta \approx 16$ they find a critical threshold of separation, which is somewhat comparable to our separation of 2 and 5 (equivalent to a $\delta$ of 4 and 7, where we see a changing fate of the array of clouds.}

This implies that these previous multi-cloud studies are not readily comparable to the newer individual cloud-crushing simulations that focus on the survival criterion and the regime of mass growth, which we addressed in this study.
While vice-versa we did not focus on the rate of destruction, we do, however, confirm findings of previous work that found prolonged survival because of more or closer spaced clouds (cf., e.g., Fig. \ref{fig:series1_mass} and Fig. \ref{fig:series1_dist_mass}). In agreement with single-cloud studies we also find, however, that the main decisive factor for the cold mass evolution is the strength of cooling. For instance, only changing the cooling efficiency by a factor $<2$ can change the cold gas mass at a fixed time by orders of magnitude whereas changing the `shielding' in adiabatic runs affects the fraction of survival mostly by a factor of a few\footnote{As pointed out in \S~\ref{res:spheres-random}, the `shielding' with cooling leads to additional non-linear effects to the increased amount of mixed gas. This furthermore complicates the comparison with previous work.}.

\subsection{Caveats and outlook}
We neglect important physical effects that may limit the applicability of our results to observed systems. 
\begin{itemize}
\item Substantive magnetic fields extend kpc from galactic midplanes \citep[e.g., see][for a recent review of the CHANG-ES survey]{irwin2024chang} and can significantly modify the survival condition of cold clouds in hot winds. Specifically, recent studies by \citet{hidalgo2024better,kaul2025tales} find that the `survival criterion' discussed in this work is shifted by orders of magnitude (for a Plasma beta $\beta\sim 1$ wind) due to a combination of a lowering of the overdensity and an enhanced acceleration due to magnetic draping \citep{dursi2008draping,McCourt_OLeary_Madigan_Quataert_2015}. We expect that magnetic fields can influence the multi-cloud criterion presented here significantly. Due to the additional compression, single cloud runs with magnetic fields generally find a decreased extent of the mixed gas \citep[e.g.][]{Gronke_Oh_2020} which might lead to a smaller surrounding `box' compared to the here presented hydrodynamical runs. Furthermore, cloud tails tend to be significantly extended, yet since they tend to be quite diffuse, it remains unclear to what extent this affect may change our findings on the importance of relatively small inter-cloud distances. In both cases, this would need to be tested with additional simulations.
\item Similarly, cosmic rays might even dominate the pressure of the circumgalactic medium \citep[see the multitude of references within theses recent reviews][]{owen2023cosmic,ruszkowski2023cosmic}; cosmic rays may accelerate the entrainment process, depending on their transport \citep[][]{bruggen2020launching,bustard2021cosmic,huang2022cosmic,weber2025crexit}. Thus far no quantitative study of cosmic ray acceleration of efficiently cooling clouds has been conducted, but we expect that a similar `survival criterion' for a multi-cloud configuration may exist \citep[importantly, with altered cooling efficiency due to cosmic rays;][]{Butsky2020}.
\item Furthermore, we do not simulate the galactic context of these clouds, which is important when comparing to observed high velocity clouds in the Galaxy \citep[][]{jung2023magnetic,zhang2025starburst}. While we expect that the change in the background primarily affects the enclosing volume through $r_{\rm crit}$, which is taken into account here, this would need to be explicitly tested.
\item Microscopic effects such as thermal conduction and viscosity have also not been considered in this work. However, while in principle they can affect the cooling efficiency, we note that for both for conduction (\citealp{tan2021radiative}, see also the discussion in \citealp{fielding2022structure}) and viscosity \citep{MarinGilabert2025} it has been shown that in the regime we consider the mass transfer rate is unchanged. 
Larger scale `cloud crushing' simulations that include thermal conduction and / or viscosity \citep{Bruggen2016,Bruggen2023,Li_Hopkins_Squire_Hummels_2020} show, however, an altered morphology which may affect the `enclosing volume' parameters discussed in this work.
\item Stochasticity in the evolution of the systems also plays an important effect, especially for configurations, like the ellipsoidal, where the random draw of a set of clouds that make up the enclosing shape can affect shielding between clouds and significantly alter the volume of influence for the individual clouds.
Multiple simulations of similar shapes can counter these effects, and the distinction between survival and destruction may be phrased as a statistical statement averaging over the individual runs. 
Here, we did not quantify the effect of stochasticity which would require performing several simulations with the same large scale run parameters. We do expect the effect of stochasticity to play a role close to the critical volume filling fraction and to be less important deep inside each regime \citep[akin to the one found for turbulent media][]{Gronke2022}.
\end{itemize}

Nevertheless, our work builds on previous efforts to determine the survival of clouds under hydrodynamic conditions. 
The multi-cloud survival criterion appears to be a reliable indicator for survival and destruction of complex morphology.
In future work, we aim to study the application of our framework to realistic configurations of an ensemble of clouds possibly ranging towards a ``no-cloud'' approach where we consider density distribution functions in the ISM (finding a comparable measure of \textit{cloud size} will be a key hurdle when applying the effective volume filling fraction).

\section{Conclusion} \label{sec:conclusion}
In this work, we perform three-dimensional hydrodynamic cloud crushing simulations with radiative cooling to specifically study how multiple clouds impact the survival of the cold gas. We systematically vary the number of clouds, the inter-cloud separation, and the geometric placement of the clouds and find a unified multi-cloud survival criterion based on an \textit{effective volume filling fraction} related to the single-cloud survival criterion. 
Summarized we find (i) increasing cloud number in wind-aligned configurations transitions from rapid destruction to sustained growth, driven by tail merging and mutual shielding; (ii) orthogonally arranged clouds experience severe mass loss unless periodic conditions introduce downstream mass replenishment; (iii) survival probabilities sharply decrease once inter-cloud spacing exceeds a few cloud radii, isolating clouds and suppressing turbulent mass feeding; and (iv) highly fragmented cloud ensembles exhibit enhanced resilience compared to fewer larger clouds, underscoring the role of turbulent mixing layers.

Our key findings in detail are as follows:
\begin{itemize}
    \item For wind-aligned placements of clouds, sufficiently large inter-cloud distances render the survival of cold gas equivalent to the single cloud case. However, once the inter-cloud distance is reduced to the point that clouds' tails merge, then a sufficiently large number of clouds enables survival. The precise number of clouds required for survival decreases (increases) with inter-cloud distance (for decreasing radiative cooling efficiency). At zero inter-cloud distance (i.e., clouds intersecting at only one point), 35 clouds are required for survival for $\ratioinline \sim 1$ whereas $\sim$60 clouds are needed for survival with $\ratioinline \sim 1.6$. 
    \item When placing clouds in a single row orthogonal to the wind, every case we simulated led to destruction, or borderline survival in the case of 60 clouds for $\ratioinline \sim 1$. It is possible an even larger number of clouds would lead to survival as the 60 cloud case suggests the tails mix downstream and might form a persistent site for cold gas growth above some mass threshold, which we leave for future work. 
    \item For cases with an infinite band of clouds orthogonal to the wind (achieved using periodic boundary conditions), we find the cold gas can survive if a critical number of rows of clouds is achieved. The precise number of rows increases with the inter-cloud distance. For an inter-cloud distance of 10 $r_{\rm cloud}$, even the largest number of rows we considered (20) showed swift destruction.
    \item For a more realistic 3D placement of clouds within an enclosing volume that is ellipsoidal (including spherical, semi-major axis that is wind-aligned and wind-orthogonal), we find a sufficiently large number of clouds enables survival. However, our results suggest that, surprisingly, there does not exist a simple volume filling fraction threshold for survival.
    \item Consistent with our findings that decreasing the inter-cloud separation promotes cold phase survival and mass growth, we find that, at fixed volume fraction and $\ratioinline \sim 1$ for the individual clouds, a smaller number of large clouds suffers an enhanced period of mass loss compared to a larger number of small clouds randomly placed within an ellipsoidal 3D enclosing volume.
\end{itemize}

Using these numerical results, we find an effective volume filling fraction that reliably separates surviving and destroyed systems and has a critical threshold motivated by both the single cloud survival criterion and considerations of the geometric configuration of the system (cf. \S~\ref{sec:survival_criterion}).
    Besides some stochastic effects in our cloud configurations that affect the tail formation and mass-feeding mechanisms, various configurations are captured by a single value $F_V$.
    This finding can be treated as a general multi-cloud survival criterion that may be applied to an arbitrarily complex cloud morphology if the concept of filling fraction can be employed there.

In summary, we have shown that various geometries can be compared through a single volume filling fraction and reliably classified through a single multi-cloud survival criterion.
It remains for future work to expand the parameter space in the dimensions of $\ratioinline$, $\chi$, and $\Mwind$ to verify the validity of the newly found survival criterion.
While this study focused on purely hydrodynamic setups, future work will need to explore the impact of magnetic fields and other physical processes that may modify the proposed survival criterion. Likewise, implementing more realistic initial conditions with complex, density structures -- as expected in galactic environments -- will be essential for extending the applicability of our findings. Despite these challenges, the multi-cloud survival criterion introduced here offers a promising framework to assess cold gas survival in a broad range of astrophysical settings, and paves the way for more comprehensive studies of multiphase interactions in realistic galactic outflows.

\section*{Acknowledgments}
We want to thank Hitesh Kishore Das and Fernando Hidalgo Pineda for helpful discussions.
Computations were performed on the HPC system Freya at the Max Planck Computing and Data Facility.
MG thanks the Max Planck Society for support through the MPRG.
BS thanks the Max Planck Institute for Astrophysics for providing the opportunity and resources to conduct a bachelor thesis project, which laid the foundation for this paper during which the majority of the work was carried out.
KD acknowledges support by the COMPLEX project from the European Research Council (ERC) under the European Union’s Horizon 2020 research and innovation program grant agreement ERC-2019-AdG 882679.

This project made use of the Athena 4.2 code \citep{Stone_2008}, the yt visualization toolkit \citep{2011ApJS..192....9T,turk2024introducing}, the Matplotlib Style \textit{SciencePlots} \citep{SciencePlots}, CMasher \citep{Cmasher}, the IPython package \citep{PER-GRA:2007}, NumPy \citep{harris2020array}, SciPy \citep{2020SciPy-NMeth}, and matplotlib, a Python library for publication quality graphics \citep{Hunter:2007}.


\section*{Data Availability}
The data underlying this paper will be shared on reasonable request to the corresponding author.


\bibliographystyle{mnras}
\bibliography{bibliography}



\FloatBarrier
\appendix
\section{Cuboid-shaped Boxes in the Effective Volume Filling Fraction} \label{app:cuboid_fv}

\begin{figure*}
    \centering
    \includegraphics[width=\linewidth]{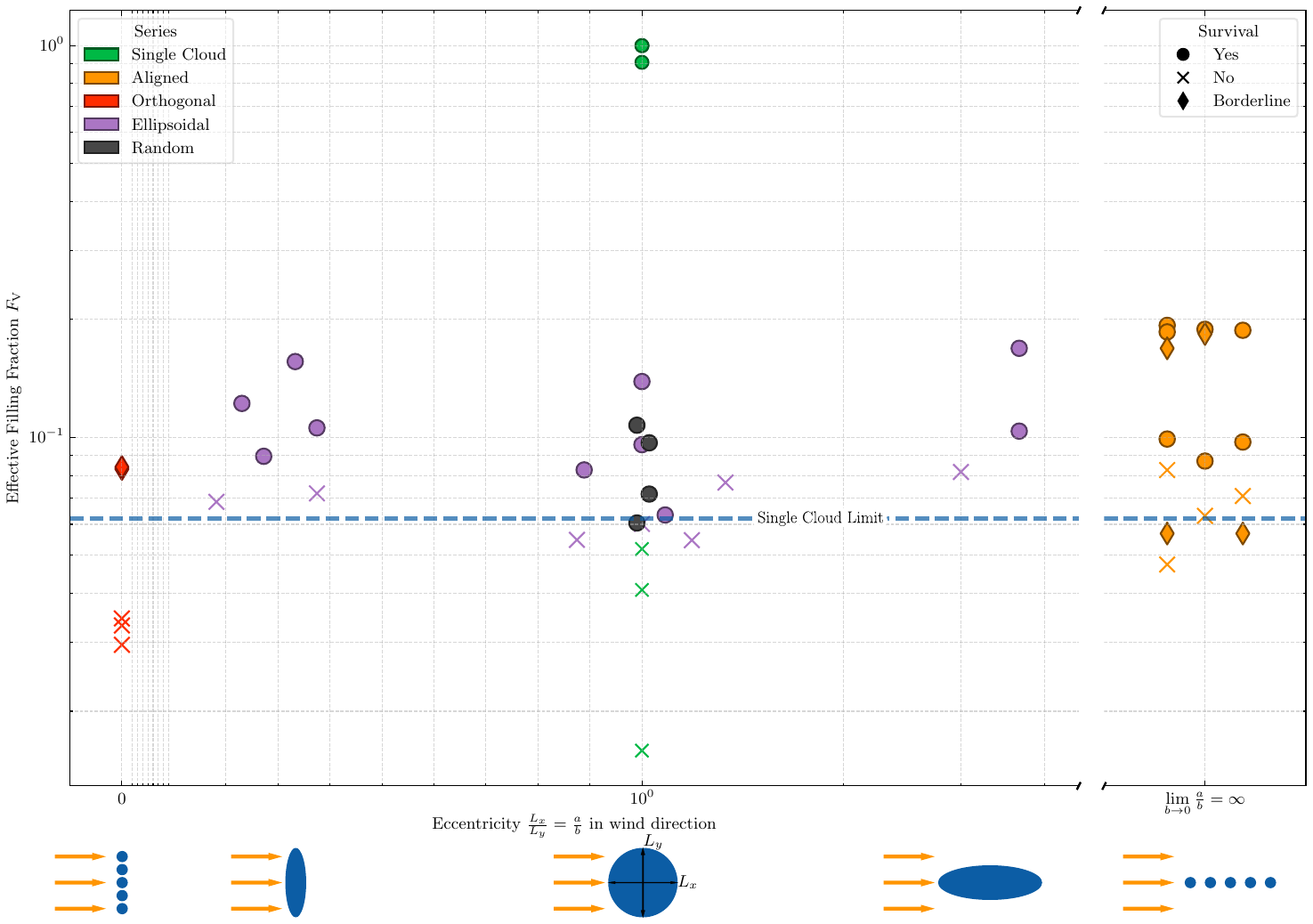}
    \caption{Overview of effective volume filling fraction based on a cuboid box (width $L_\perp=\alpha\rcrit=3\rcrit$, length $L_\parallel=\beta\rcrit=7.5\rcrit$). The eccentricity is calculated as in \autoref{fig:unified_cone} and $F_V$ is cut off at a value of 1. The critical threshold $F_{V,\mathrm{crit}}=4/3\pi\alpha^{-2}\beta^{-1}$ shows a separation in the $F_V$ dimension, that is however not as clear as in the cuboid case.}
    \label{fig:unified_cuboid}
\end{figure*}

Analogous to the conic box in \autoref{sec:survival_criterion} we also probe a cuboid box around each cloud. The process is identical, but neighboring clouds, especially in the wind aligned and wind-orthogonal configurations will show the biggest impact (more overlap between neighboring clouds, thus larger volume filling fraction compared to ellipsoidal setups).
\autoref{app:cuboid_fv} shows the results.
We find that no surviving system falls below the single cloud limit, but on the other side, many destroyed systems fall above the line by 0.1 to 0.2 in $F_V$.
The changes show especially in both aligned and orthogonal systems (high wind area ellipsoidal), where many systems that fell below the line for a conic box now appear above the critical threshold.
This shows that a cuboid box fails to capture the morphologies in these extreme cases, because it rewards configurations where clouds are close together in the wind orthogonal direction (like the high-wind area ellipsoidal setup).
These, however, are configurations more exposed to the wind and less likely to survive. 

\section{YZ-Periodic Patterns} \label{app:periodic}

\begin{figure*}
    \centering
    \includegraphics[width=\textwidth]{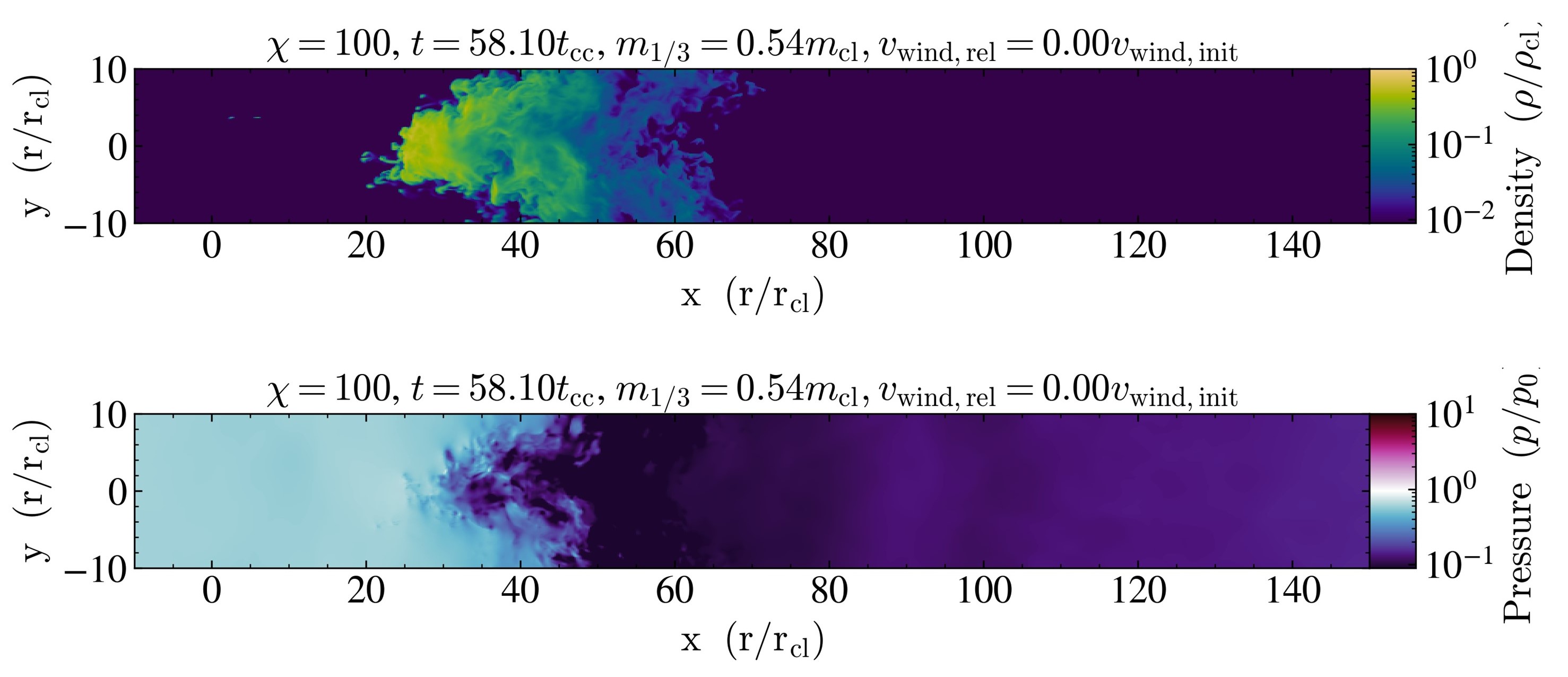}
    \caption{A snapshot of a yz-periodic simulation.
The pressure slice shows a region of low pressure, which leads to altered cooling.
This effect is not a physical occurrence and is caused by the boundary effects of the periodic box.
The time of the snapshot can be directly linked to either extreme fluctuations in the mass growth figure, but also with a sharp turning edge in mass growth for earlier times ($\approx 20\, \tcc$ in the simulations).}
    \label{fig:series2_pressureartefacts}
\end{figure*}

\begin{figure}
    \centering
    \includegraphics[width=\columnwidth]{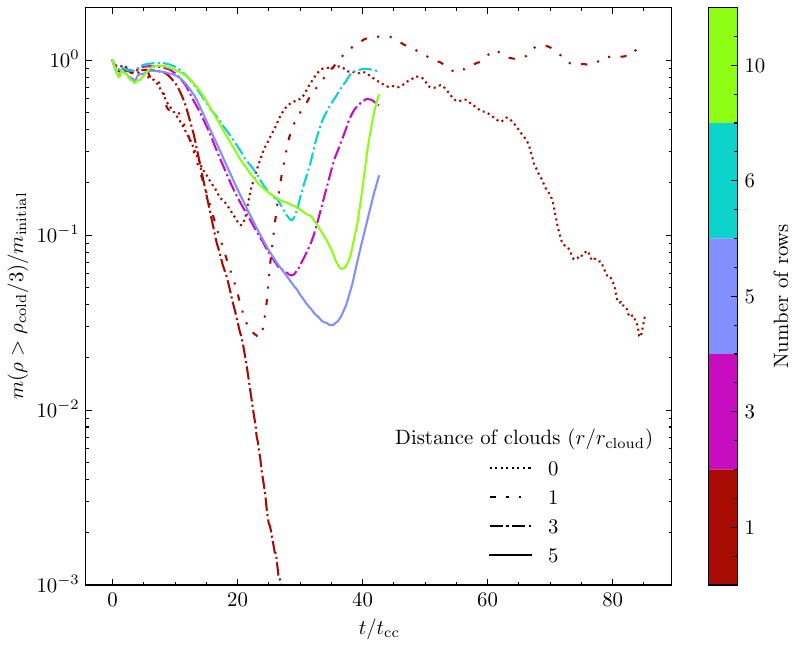}
    \caption{The relative mass growth for a simulation periodic in y and z direction is being displayed up to 80 cloud crushing times and for a varying number of rows and edge-to-edge distance.
The overall setup now looks like an infinite slab of clouds facing the wind.
The rows are being added in the direction of the wind, causing the slab to slowly turn into a cube-shaped setup within the periodic box.
The general trends hold -- more gas mass that is initially placed downwind of the clouds that experience first impact with the wind supports survival.
Larger separation generally suppresses cloud growth, but can be compensated by more rows downwind.
Most of the simulations suffer from numerical effects -- either altered cooling due to lower pressure, or cold gas mass leaving the simulation box towards the left once the box is co-moving -- imprinted in the strong fluctuation at later stages and sharp turnover points in the initial phase.}
    \label{fig:series2_yzP_mass-label}
\end{figure}

\begin{figure}
    \centering
    \includegraphics[width=\columnwidth]{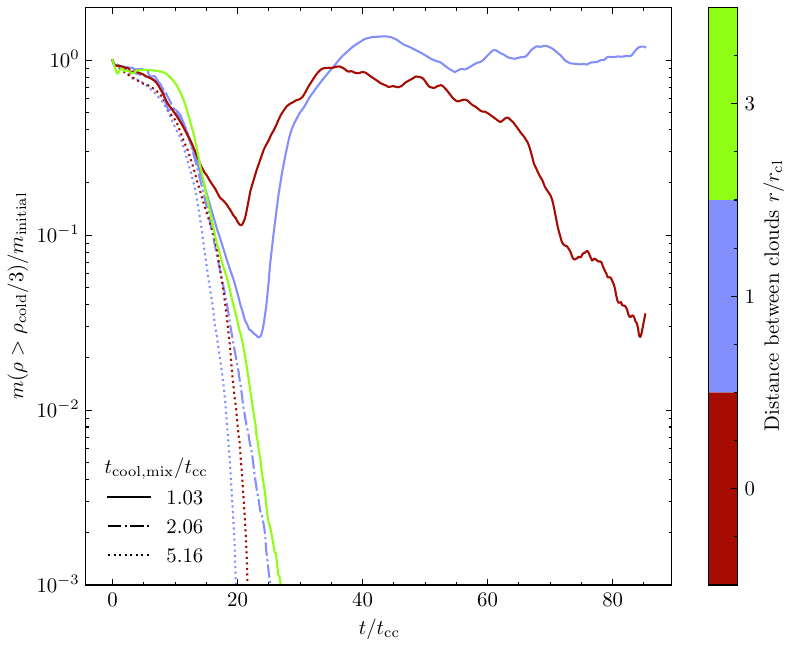}
    \caption{The mass growth evolution of cold gas mass for a setup of yz-periodically placed clouds with one row in the x-direction.
The fluctuations in the solid lines are results of low pressure areas and partly cold gas leaving the simulation box.
The general trend holds -- higher spatial separation and likewise higher ratios $\ratioinline$ of the single cloud suppress cold gas growth and thus survival.}
    \label{fig:series2_mass_ratios}
\end{figure}

To implement periodic boundaries in both the y and z directions, we extend the box in z similarly to the y-periodic setup. This involves replicating cloud rows in z, ensuring at least half an inter-cloud spacing between the final cloud and the box edge. The number of rows varies across runs (see Appendix \ref{app:parameters}), which may contribute to observed boundary effects.

However, this configuration introduces noticeable artifacts.
Extended simulations (up to $80\,\tcc$) show nonphysical mass fluctuations -- comparable to the total initial cold gas mass -- not attributable to numerical noise.
Figure \ref{fig:series2_pressureartefacts} highlights pressure drops in certain regions, disrupting cooling and inducing sharp changes in cold gas mass (e.g., at $t \sim 20\,\tcc$ in \autoref{fig:series2_yzP_mass-label}).
These artifacts appear after initial cloud growth, so early-time results remain usable, though caution is warranted at later stages.

Figure \ref{fig:series2_yzP_mass-label} presents the evolution of normalized cold gas mass for the yz-periodic runs. 
Line color indicates the number of x-direction rows (which affect dynamics), while linestyle denotes inter-cloud distance, now applied in all directions.
As in the y-periodic setup, larger separations hinder mass growth.
For instance, 1-row setups cease to grow beyond $1-3\,\rcloud$, while $\geq5$ rows maintain cold mass even at $5\,\rcloud$ spacing.

Finally, Figure \ref{fig:series2_mass_ratios} explores variations in $\ratioinline$. Even modest increases (e.g., $\ratioinline = 1.03$) significantly reduce survival, underscoring the sensitivity of the yz-periodic setup to this parameter.

\FloatBarrier

\section{Initial Parameters for the simulations}
\label{app:parameters}
Some parameters are not included in these tables (\ref{tab:params_aligned}, \ref{tab:params_orthogonal}, \ref{tab:params_ellispoidal}, and \ref{tab:params_ranspheres}), as they are the same for every run. These include $T_{\mathrm{floor}} = 1 \times 10^{4} \mathrm{K}$, $T_{\mathrm{ceil}} = 1 \times 10^{7}\mathrm{K}$, $T_{\mathrm{cool,floor}} = 1 \times 10^{4} \mathrm{K}$, $T_{\mathrm{cool,ceil}} = 0.6 \times 10^{6} \mathrm{K}$ (except for runs marked with $^\dagger$ where $T_{\mathrm{cool,ceil}} = 1 \times 10^{6} \mathrm{K}$), $\chi=100$, $\Tcloud=1\times 10^{4}\mathrm{K}$, and $\Mwind=1.5$. 
The general identifier of shape, e.g. SPHERE 14, refers to the enclosing shape and the digits tell the radius in units of $\rcloud$ (in case of spherical setups).
The \# column states how many clouds were in the simulation box. 
This also counts for the periodic runs, as one has to set up a box with a fixed number of clouds to be used in periodic simulations. 
Sep defines the inter-cloud separation from edge to edge; depending on periodicity this can mean that the separation is limited in some spatial dimensions.
Included are both the survival, destruction, or borderline (BL) evolution of the clouds, and which figures show the respective time evolution.

\begin{table*}
    \caption{Parameters for Aligned Morphology Runs}
    \begin{tabularx}{\textwidth}{XXXXXXXX}            \toprule
        $\ratioinline$    & \#    & Periodic & Sep         & XYZ Cells            & XYZ Length      & Survival & Figures                                                \\ \midrule
        $[1]$             & $[1]$ &          & $[\rcloud]$ & $[\mathrm{cells}^3]$ & $[\rcloud]$     &          &                                                        \\ \midrule
        $1.58246^\dagger$ & 5     & No       & 0           & 912 160 160          & 114.0 20.0 20.0 & No       & \ref{fig:series1_mass}                                 \\
        $1.58247^\dagger$ & 10    & No       & 0           & 1032 160 160         & 129.0 20.0 20.0 & No       & \ref{fig:series1_mass}                                 \\
        $1.58248^\dagger$ & 15    & No       & 0           & 1152 160 160         & 144.0 20.0 20.0 & No       & \ref{fig:series1_mass}                                 \\
        1.58248           & 35    & No       & 0           & 1104 320 320         & 138.0 40.0 40.0 & No       & \ref{fig:series1_mass}                                 \\
        1.58248           & 60    & No       & 0           & 2240 160 160         & 280.0 20.0 20.0 & Yes      & \ref{fig:series1_mass}                                 \\
        1.03277           & 15    & No       & 0           & 1152 160 160         & 144.0 20.0 20.0 & BL       & \ref{fig:series1_mass}                                 \\
        1.03277           & 30    & No       & 0           & 1032 320 320         & 129.0 40.0 40.0 & BL       & \ref{fig:series1_mass},    \ref{fig:series1_dist_mass} \\
        1.03277           & 35    & No       & 0           & 1104 320 320         & 138.0 40.0 40.0 & Yes      & \ref{fig:series1_mass},    \ref{fig:series1_dist_mass} \\
        1.03277           & 40    & No       & 0           & 1032 320 320         & 129.0 40.0 40.0 & Yes      & \ref{fig:series1_mass},    \ref{fig:series1_dist_mass} \\
        1.03277           & 50    & No       & 0           & 1334 320 320         & 168.0 40.0 40.0 & Yes      & \ref{fig:series1_mass},    \ref{fig:series1_dist_mass} \\
        1.03277           & 60    & No       & 0           & 2240 160 160         & 280.0 20.0 20.0 & Yes      & \ref{fig:series1_mass},    \ref{fig:series1_dist_mass} \\
        1.03277           & 35    & No       & 2           & 1680 320 320         & 206.0 40.0 40.0 & Yes      & \ref{fig:series1_dist_mass}                            \\
        1.03277           & 35    & No       & 5           & 2460 320 320         & 308.0 40.0 40.0 & BL       & \ref{fig:series1_dist_mass}                            \\
        1.03277           & 60    & No       & 2           & 2448 320 320         & 306.0 40.0 40.0 & Yes      & \ref{fig:series1_dist_mass}                            \\
        1.03277           & 60    & No       & 5           & 3860 320 320         & 482.6 40.0 40.0 & BL      & \ref{fig:series1_dist_mass}                            \\ \bottomrule
    \end{tabularx}

    \label{tab:params_aligned}
\end{table*}

\begin{table*}
    \caption{Parameters for Orthogonal Runs}
    \begin{tabularx}{\textwidth}{XXXXXp{1.8cm}p{1.8cm}XXX}
        \toprule
        $\ratioinline$    & \#    & Rows  & Periodic & Sep         & XYZ Cells            & XYZ Length       & Survival & Z-Rows & Figures                                                         \\ \midrule
        $[1]$             & $[1]$ & $[1]$ &          & $[\rcloud]$ & $[\mathrm{cells}^3]$ & $[\rcloud]$      &          &        &                                                                 \\ \midrule
        $1.58246^\dagger$ & 5     & 1     & No       & 0           & 2720~160~160         & 340.0~20.0~20.0  & No       &        & \ref{fig:series2_nonP_mass}                                     \\
        $1.58246^\dagger$ & 10    & 1     & No       & 0           & 2720~240~240         & 340.0~30.0~30.0  & No       &        & \ref{fig:series2_nonP_mass}                                     \\
        $1.58246^\dagger$ & 15    & 1     & No       & 0           & 2720~368~368         & 340.0~46.0~46.0  & No       &        & \ref{fig:series2_nonP_mass}                                     \\
        $1.03277^\dagger$ & 40    & 1     & No       & 0           & 1040~800~160         & 130.0~100.0~20.0 & BL       &        & \ref{fig:series2_nonP_mass}                                     \\
        1.03277           & 60    & 1     & No       & 0           & 1040~1120~160        & 130.0~140.0~20.0 & BL       &        & \ref{fig:series2_nonP_mass}                                     \\ \midrule
        1.03277           & 20    & 2     & Y        & 1           & 720~240~160          & 90.0~30.0~20.0   & BL       & 1      & \ref{fig:series2_yP_mass}                                       \\
        1.03277           & 20    & 2     & Y        & 5           & 752~560~160          & 94.0~70.0~20.0   & No       & 1      & \ref{fig:series2_yP_mass}                                       \\
        1.03277           & 30    & 3     & Y        & 1           & 768~240~160          & 96.0~30.0~20.0   & Yes      & 1      & \ref{fig:series2_yP_mass}                                       \\
        1.03277           & 30    & 3     & Y        & 5           & 792~560~160          & 99.0~70.0~20.0   & No       & 1      & \ref{fig:series2_yP_mass}                                       \\
        1.03277           & 40    & 4     & Y        & 1           & 792~240~160          & 99.0~30.0~20.0   & Yes      & 1      & \ref{fig:series2_yP_mass}                                       \\
        1.03277           & 40    & 4     & Y        & 5           & 848~560~160          & 106.0~70.0~20.0  & No       & 1      & \ref{fig:series2_yP_mass}                                       \\
        1.03277           & 100   & 10    & Y        & 5           & 1224~560~320         & 153.0~70.0~40.0  & Yes      & 1      & \ref{fig:series2_yP_mass}                                       \\
        1.03277           & 200   & 20    & Y        & 10          & 2544~950~320         & 318.0~118.8~40.0 & No       & 1      & \ref{fig:series2_yP_mass}                                       \\ \midrule
        2.06553           & 9     & 1     & YZ       & 1           & 1280~72~72           & 160.0~9.0~9.0    & No       & 3      & \ref{fig:series2_mass_ratios}                                   \\
        5.16383           & 9     & 1     & YZ       & 1           & 1281~72~72           & 160.2~9.0~9.0    & No       & 3      & \ref{fig:series2_mass_ratios}                                   \\
        5.16383           & 5     & 1     & YZ       & 0           & 1280~80~16           & 160.0~10.0~2.0   & No       &        & \ref{fig:series2_mass_ratios}                                   \\
        1.03277           & 30    & 1     & YZ       & 0           & 1280~160~48          & 160.0~20.0~6.0   & Yes      & 3      & \ref{fig:series2_yzP_mass-label}, \ref{fig:series2_mass_ratios} \\
        1.03277           & 20    & 1     & YZ       & 1           & 1280~240~48          & 160.0~30.0~6.0   & Yes      & 2      & \ref{fig:series2_yzP_mass-label}, \ref{fig:series2_mass_ratios} \\
        1.03277           & 10    & 1     & YZ       & 3           & 1280~400~40          & 160.0~50.0~5.0   & No       & 1      & \ref{fig:series2_yzP_mass-label}, \ref{fig:series2_mass_ratios} \\
        1.03277           & 30    & 3     & YZ       & 3           & 800~400~40           & 100.0~50.0~5.0   & Yes      & 1      & \ref{fig:series2_yzP_mass-label}                                \\
        1.03277           & 50    & 5     & YZ       & 5           & 944~560~35           & 118.0~70.0~11.2  & Yes      &        & \ref{fig:series2_yzP_mass-label}                                \\
        1.03277           & 60    & 6     & YZ       & 3           & 920~400~120          & 115.0~50.0~5.0   & Yes      &        & \ref{fig:series2_yzP_mass-label}                                \\
        1.03277           & 100   & 10    & YZ       & 5           & 1224~560~35          & 153.0~70.0~7.0   & Yes      & 1      & \ref{fig:series2_yzP_mass-label}                                \\ \bottomrule
    \end{tabularx}

    \label{tab:params_orthogonal}
\end{table*}

\begin{table*}
    \caption{Parameters for Ellipsoidal Runs}
    \begin{tabularx}{\textwidth}{XXXXXXX}
        \toprule
        $\ratioinline$ & \#    & Shape      & XYZ Cells            & XYZ Length      & Survival & Figures                                                         \\ \midrule
        $[1]$          & $[1]$ &            & $[\mathrm{cells}^3]$ & $[\rcloud]$     &          &                                                                 \\ \midrule
        1.03277        & 15    & SPHERE 6   & 1000 560 560         & 125.0 70.0 70.0 & No       & \ref{fig:series3_spheres_mass}                                  \\
        1.03277        & 32    & SPHERE 6   & 1000 560 560         & 130.0 70.0 70.0 & Yes      & \ref{fig:series3_spheres_mass}, \ref{fig:series4_massevolution} \\
        1.03277        & 64    & SPHERE 6   & 1280 640 640         & 160.0 80.0 80.0 & Yes      & \ref{fig:series3_spheres_mass}                                  \\
        1.03277        & 15    & SPHERE 10  & 1000 560 560         & 130.0 70.0 70.0 & No       & \ref{fig:series3_spheres_mass}                                  \\
        1.03277        & 32    & SPHERE 10  & 1000 560 560         & 130.0 70.0 70.0 & No       & \ref{fig:series3_spheres_mass}                                  \\
        1.03277        & 64    & SPHERE 10  & 1280 640 640         & 160.0 80.0 80.0 & Yes      & \ref{fig:series3_spheres_mass}                                  \\
        1.03278        & 64    & SPHERE 14  & 1441 640 640         & 180.2 80.0 80.0 & Yes      & \ref{fig:series3_spheres_mass}                                  \\
        1.03277        & 64    & SPHERE 20  & 1440 640 640         & 180.0 80.0 80.0 & No       & \ref{fig:series3_spheres_mass}                                  \\ \midrule
        1.03277        & 15    & ELL HIGH   & 1080 480 480         & 125.0 60.0 60.0 & No       & \ref{fig:series3_ellips_mass}                                   \\
        1.03277        & 30    & ELL HIGH   & 1080 480 480         & 125.0 60.0 60.0 & Yes      & \ref{fig:series3_ellips_mass}                                   \\
        1.03277        & 64    & ELL HIGH   & 1000 480 480         & 125.0 60.0 60.0 & Yes      & \ref{fig:series3_ellips_mass}                                   \\
        1.03277        & 15    & ELL HIGHER & 1000 480 480         & 125.0 60.0 60.0 & No       & \ref{fig:series3_ellips_mass}                                   \\
        1.03277        & 30    & ELL HIGHER & 1000 480 480         & 125.0 60.0 60.0 & Yes      & \ref{fig:series3_ellips_mass}                                   \\
        1.03277        & 64    & ELL HIGHER & 1000 480 480         & 125.0 60.0 60.0 & Yes      & \ref{fig:series3_ellips_mass}                                   \\ \midrule
        1.03277        & 15    & ELL LOW    & 1000 560 560         & 130.0 70.0 70.0 & No       & \ref{fig:series3_ellips_mass}                                   \\ \
        1.03277        & 32    & ELL LOW    & 1000 560 560         & 130.0 70.0 70.0 & Yes      & \ref{fig:series3_ellips_mass}                                   \\
        1.03277        & 64    & ELL LOW    & 1280 640 640         & 160.0 80.0 80.0 & Yes      & \ref{fig:series3_ellips_mass}                                   \\ \bottomrule
    \end{tabularx}

    \label{tab:params_ellispoidal}
\end{table*}

\begin{table*}
    \caption{Parameters for Random Spherical Runs}
    \begin{tabularx}{\textwidth}{XXXXXXX}
        \toprule
        $\ratioinline$ & \#    & Shape        & XYZ Cells            & XYZ Length       & Survival & Figures                         \\ \midrule
        $[1]$          & $[1]$ &              & $[\mathrm{cells}^3]$ & $[\rcloud]$      &          &                                 \\ \midrule
        1.72128        & 150   & RANSPHERE 30 & 1840 773 773         & 230.0 96.7 96.7  & Yes      & \ref{fig:series4_massevolution} \\
        1.03494        & 150   & RANSPHERE 30 & 1841 773 773         & 230.0 96.7 96.7  & Yes      & \ref{fig:series4_massevolution} \\
        1.29096        & 64    & RANSPHERE 30 & 1350 580 580         & 172.5 72.5 72.5  & Yes      & \ref{fig:series4_massevolution} \\
        1.03277        & 64    & RANSPHERE 30 & 1351 580 580         & 172.75 72.5 72.5 & Yes      & \ref{fig:series4_massevolution} \\\bottomrule
    \end{tabularx}
    \label{tab:params_ranspheres}
\end{table*}
\bsp	
\label{lastpage}
\end{document}